\documentclass[a4paper,fleqn,usenatbib,useAMS]{mnras}
\usepackage[T1]{fontenc} 
\usepackage[UKenglish]{babel} 
\usepackage{lmodern} 

\usepackage{amsfonts,amssymb,amsmath}
\usepackage{amsbsy,color}

\usepackage{graphicx}
\graphicspath{{"PATH FOR FIGURES HERE"}} 

\usepackage[absolute]{textpos} 

\DeclareMathOperator*{\argmax}{arg\,max}
\DeclareMathOperator*{\argmin}{arg\,min}

\newcommand{\Btest}{$\mathcal{B}_\text{test}$}
\newcommand{\Btrain}{$\mathcal{B}_\text{train}$}
\newcommand{\Utest}{$\mathcal{U}_\text{test}$}
\newcommand{\Utrain}{$\mathcal{U}_\text{train}$}
\newcommand{\E}{{\rm E}}
\newcommand{\Var}{{\rm Var}}
\newcommand{\fobs}{f_{\rm obs}}
\newcommand{\GP}{{\rm GP}}
\newcommand{\band}{c}
\newcommand{\bandset}{C}
\newcommand{\Data}{{\cal Y}}

\newcommand{\customwidth}{0.5\textwidth}
\newcommand{\rt}[1]{\textcolor{red}{\bf{[RT: #1]}}}
\newcommand{\ear}[1]{\textcolor{green}{\bf{[EAR: #1]}}}

\newcommand{\SN}{SNIa}
\newcommand{\bias}[1]{$\mathcal{B}_{#1}$}
\newcommand{\unbias}[1]{$\mathcal{U}_{#1}$}

\newcommand{\edit}[1]{{\color{blue}{#1}}}

\begin{document}
\setlength{\unitlength}{1mm}
\title[STACCATO]{STACCATO: A Novel Solution to Supernova Photometric Classification with Biased Training sets}

\author[Revsbech, Trotta \& van Dyk]{Esben A. Revsbech$^{1}$, R.~Trotta$^{2,3,4}$\thanks{Corresponding author:
r.trotta@imperial.ac.uk} and David~A.~van Dyk$^{1,3,4}$\\
$^{1}$Statistics Section, Mathematics Department,  Huxley Building, South Kensington Campus, Imperial College London,  London SW7 2AZ\\
$^{2}$Astrophysics Group, Physics Department, Imperial College London, Prince Consort Rd, London SW7 2AZ\\
$^{3}$Imperial Centre for Inference and Cosmology, Astrophysics Group, Blackett Laboratory, Prince Consort Rd, London SW7 2AZ \\
$^{4}$Data Science Institute, William Penney Laboratory, Imperial College London, London SW7 2AZ
}
\date{\today}

\maketitle

\begin{abstract}
We present a new solution to the problem of classifying Type~Ia supernovae from their light curves alone given a spectroscopically confirmed but biased training set, circumventing the need to obtain an observationally expensive unbiased training set. We use Gaussian processes (GPs) to model the supernovae's (SN) light curves, and demonstrate that the choice of covariance function has only a small influence on the GPs ability to accurately classify SNe. We extend and improve the approach of \citet{Richards2012} --- a diffusion map combined with a random forest classifier --- to deal specifically with the case of biassed training sets. We propose a novel method, called STACCATO (`\emph{S}yn\emph{t}hetically \emph{A}ugmented Light \emph{C}urve \emph{C}lassific\emph{at}i\emph{o}n') that synthetically augments a biased training set by generating additional training data from the fitted GPs. Key to the success of the method is the partitioning of the observations into subgroups based on their propensity score of being included in the training set. Using simulated light curve data, we show that STACCATO increases performance, as measured by the area under the Receiver Operating Characteristic curve (AUC), from 0.93 to 0.96, close to the AUC of 0.977 obtained using the `gold standard' of an unbiased training set and significantly improving on the previous best result of 0.88. STACCATO also increases the true positive rate for SNIa classification by up to a factor of 50 for high-redshift/low brightness SNe.

\end{abstract}
\bigskip

\maketitle
%

\section{Introduction}\label{ch:introduction}

Supernovae Type~Ia (\SN) have been crucial in establishing the accelerated expansion of the Universe \citep{perlmutter1999,riess1998}. They are expected to remain important distance indicators in the next few years, as the worldwide sample of \SN\ is set to increase many fold thanks to ongoing and upcoming observational programmes. The Dark Energy Survey (DES) is expected to observe $\sim 3,000$ \SN\ over 5 years, while the Large Synoptic Survery Telescope (LSST) is expected to observe of the order of $\sim 10^5$ \SN\ each year \citep{2009arXiv0912.0201L}.

One of the major bottlenecks for the cosmological analysis of such a large and rich data set is the classification of \SN\ candidates. Traditionally, \SN\ candidates have been confirmed by spectroscopic follow-up, as \SN\ are characterized by the lack of H in their spectrum and the presence of strong SiII lines. However, spectroscopic follow up of thousands of candidates is simply not observationally feasible. Thus it is becoming crucial to reliably classify \SN\ on the basis of photometric information alone~\citep{kessler2010}.  In parallel, methods are being developed to fit cosmological models to a SN sample contaminated by non-type Ia~\citep{Kunz:2006ik,Hlozek:2011wq,Knights:2012if,Kessler:2016uwi}, which generally require as  input the probability for a given object to be of type Ia (based on photometry alone).

To address this problem, \citet{kessler2010} set up a ``Supernova photometric classification challenge'', inviting teams to develop methods for \SN\ classification from a suite of numerical simulations designed to mock data from DES. The simulations contain SN Type~Ia, Ib, Ic and II light curves (LCs) with realistic noise and selection effects and are divided into a training set and a testing set. SN types are known in the training set so that it can be used to tune the classifier. During the challenge, SN types for the test set were not revealed until completion of the challenge so that this set could be used to evaluate the performance of the proposed classifiers.  Teams were evaluated on a Figure of Merit (FoM) that combines both the efficiency and the purity of the classification of the test set. There were 13 entries from 10 teams, using a variety of strategies. The broad conclusions from the original challenge are that none of the methods adopted was clearly superior, and that an overall major difficulty is the fact that the realistic training set was not representative of the test set. This is a consequence of the fact that the observer-frame magnitude cut used to define the training set map onto different brightness levels as a function of redshift for SNIa and core collapse SNe (as their rest-frame spectrum is different). As a consequence, the ratio of SNIa to non-Ia SNe is different for the training and the test set, as they have different magnitude cuts, with the proportion of non-Ia SNe being generally underrepresented at high redshift/low apparent brightness. Moreover, there are very few  high redshift or low apparent magnitude SNe \emph{of any type} in the training set. Unfortunately, tuning a classifier with such an unrepresentative training set leads to poor results, especially for high redshift and/or dim SNe.  \citet{Newling2011,Varughese2015,Lochner2016} found considerable decreases in performance if the training set is biased. \citet{Richards2012} suggested and evaluated several strategies for choosing additional \SN\ for spectroscopic follow-up. \citet{Lochner2016} carried out a comparison of five different methods and found that the classification of all of them degraded significantly when using a biased training set (of the kind that is likely to be available in practice) as opposed to a representative sample. They concluded that a representative training set is ``essential'' for good classification when using these methods.

Accurate classification based on a biased training set is generally problematic. In this work we build on the methods of \citet{Richards2012} and \citet{Lochner2016}, and suggest a novel approach to \SN\ photometric classification that only uses a biased training set. We name this classification approach, `\emph{S}yn\emph{t}hetically \emph{A}ugmented Light \emph{C}urve \emph{C}lassific\emph{at}i\emph{o}n' (STACCATO)\footnote{An R code implementing STACCATO and allowing the user to reproduce all the results and figures in this paper is available from: \texttt{https://github.com/rtrotta/STACCATO}.}. In our approach, the effects of the bias are mitigated by dividing the training set into a number of groups based on the propensity score, i.e. the probability of belonging to the training set given a number of covariates (here, redshift and apparent brightness). Because some of these groups have very small training sets, we propose to augment the training set by sampling new synthetic LCs from Gaussian Processes (GPs) fit to the original training set LCs.  We show that this strategy improves the classification accuracy considerably, to a level that compares favourably with the performance obtained from a unbiased training set. In the current implementation the choice of degree of augmentation of the training sets by synthetic LCs involves an optimization step that requires knowledge of the SNe types in a subset of the test set. We leave to future work designing and demonstrating a procedure that does not require such data, which would of course not be available in a real situation.  At the end of this paper,  however, we are able to demonstrate that even without the optimization step, our augmentation strategy performs better than using the original training set without augmentation.

Our overall strategy is to use GPs to build probabilistic models for the individual LCs. We then carry out a classification step in which a diffusion map and a random forest classifier are applied. Here we draw on the work of \citet{Richards2012}, but we use a different specification of the LCs metric, and compute the diffusion map separately for each filter and only on the relevant training data. In the final, entirely novel step, we apply a propensity score to the training set, which we then augment with synthetic LCs generated probabilistically from the fitted GPs. We demonstrate that this approach circumvents the need to design an observationally expensive unbiased training set, and that the performance of the classifier (as measured by the Area under the ROC Curve, AUC) is improved from 0.92 to 0.96 (as compared to 0.977 for an unbiased training set), with most of the improvement coming from the highest redshift SN group (with the fewest SNe in its training set). This compares favourably with the best methods in \citet{Lochner2016}, which  delivered about 0.88 for the biased training set.

The structure of this paper is as follows. In Section \ref{sec:GP} we describe our Gaussian Process fit to the LC data; in Section~\ref{sec:norm} we describe a normalization and time-alignment procedure for the LCs; in Section~\ref{sec:classification} we explain how we classify SN with a diffusion map coupled with a random forest classifier; in Section~\ref{sec:results} we present the classification results and contrast the case of a biased and an unbiased training set; in Section \ref{sec:staccato} we present the STACCATO approach and discuss the improvements it brings to the classification result; finally, we conclude in Section~\ref{sec:conclusions}.

\section{Gaussian Process Light Curve Fit} 
\label{sec:GP}

\subsection{Light Curves Data and Training Sets}\label{sec:data}
We use the dataset that was originally released by \citet{kessler2010} as part of their ``Supernova photometric classification challenge''. Since the release of that dataset, an updated version has been made available~\citep{kessler2010b}, with various improvements and a few bug fixes. In order to compare our results with the outcome of the original challenge, we used exactly the same dataset that was available to the teams that entered the challenge. More recent works have used the updated dataset, and we will apply our methodology to this newer and more realistic dataset in a future paper. 

The dataset from \citet{kessler2010} contains photometric LCs from 18,321 simulated SNe. For each SN, LCs are given in four colour bands, $\bandset = (g,r,i, z)$, measured at unevenly spaced points in time. The observations are subject to measurement error and estimated standard deviations are given as part of the dataset. The number of observations times for the LCs ranges from 0 to 41 with a mean of 19.1 and a median of 21. We only use SNe with at least three observations in each band, thus reducing the dataset to 17,330 SNe. The challenge included both a set with host galaxy photometric redshift, and one without it. In this paper, we only analyze the data set with host galaxy redshift. \cite{Lochner2016} found that removing redshift information did not significantly degrade the classification performance. We will explore this aspect in future work.  

\begin{figure}
	\centering
	\includegraphics[width=\linewidth]{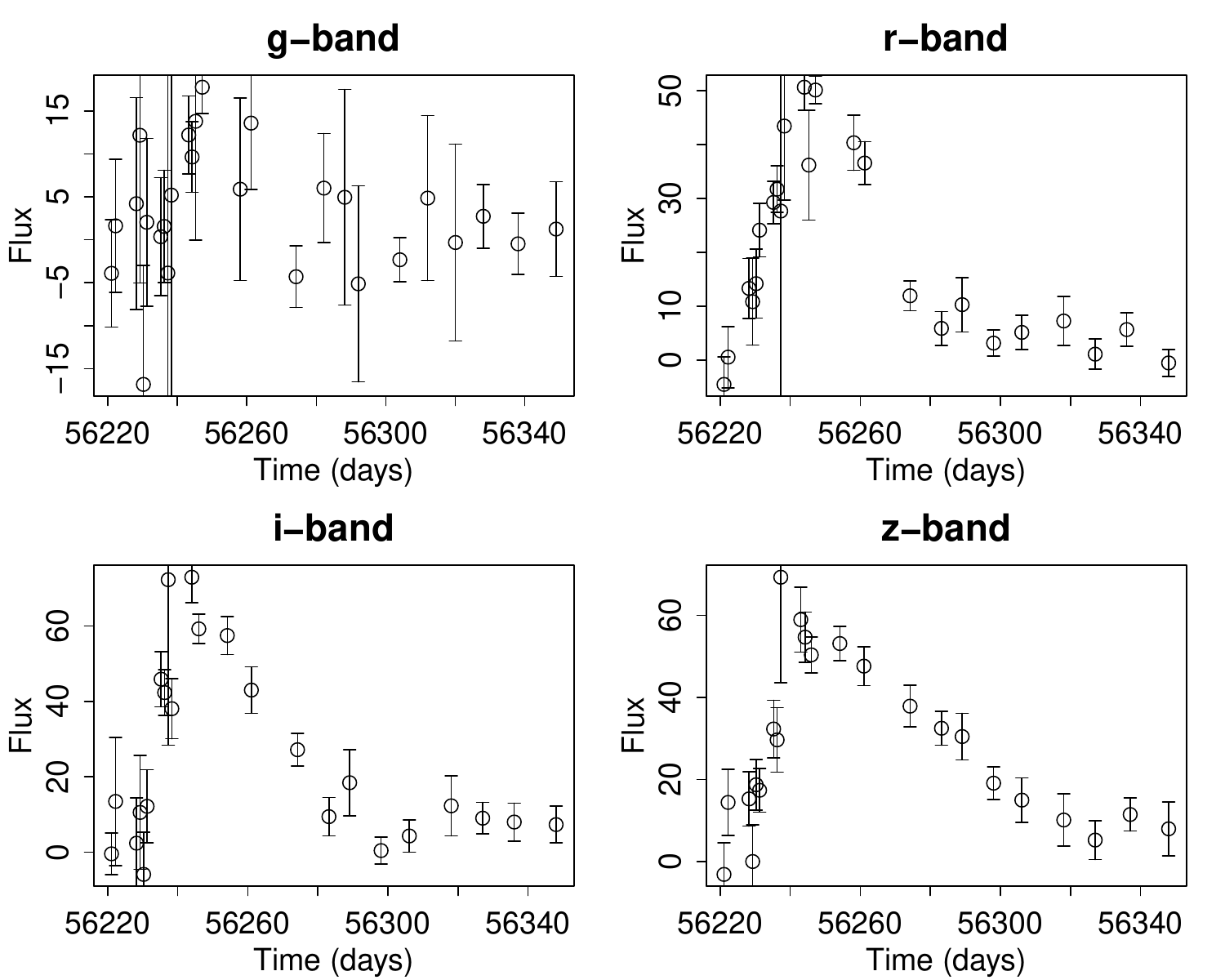}
	\caption{An example of LC data in four bands for (the randomly selected) SN194156 at $z=0.54$. Vertical $1\sigma$ error bars are also plotted.}	\label{fig1}
\end{figure}

Figure~\ref{fig1} displays the four LCs of a randomly chosen SN (at redshift~0.54)
and illustrates both the very noisy behaviour (in the $g$ band) that some SNe exhibit in one or more bands and the more `well behaved' peak structure that is typically associated with SNIa explosions.

The simulated dataset from \cite{kessler2010} is divided into a training set, \Btrain, of 1,217 SNe with known types and a test set, \Btest, of 16,113 simulated SNe with unknown types. \Btrain\ is obtained by simulating the spectroscopic follow-up efficiency from a 4m class telescope with a limiting $r$-band magnitude of 21.5, and an 8m class telescope with a limiting $i$-band magnitude of 23.5. \SN\ LCs are simulated from a mix of SALT-II and MLCS models, while non-SNIa are simulated from a set of 41 templates derived from  spectroscopically confirmed non-SNIa~\citep{kessler2010b}. The goal is to use \Btest\ to classify the SN in \Btest. Due to observational selection effects, the spectroscopic training set is biased in terms of SN types, redshift, and brightness. This bias is mimicked in the dataset of \citet{kessler2010} so that there are proportionally more bright, low redshift SNIa in \Btrain\  than in \Btest.

We also construct an {\em unbiased} training set, \Utrain, by random sampling 1,200 SNe from the entire dataset. Here we exploit the fact that the classes of the entire dataset was released post challenge. The remaining data is assigned to a corresponding test set, \Utest, used for evaluating the performance of the classifier. For consistency, the sizes of \Utrain\ and \Btrain\ are similar.  
We refer to the \Utrain\ as  `the gold standard', as it is a `best case scenario' to compare any classification algorithm against. Although, such an unbiased training set is not feasible in practice, we want to assess the reduction in the classifier performance that can be attributed to the bias in \Btrain.  The composition of both training and test sets is summarized in Table \ref{tab:datasets}. 

\begin{table*}
	\centering
	\begin{tabular}{l r l r l r l r}
		\hline \\[-1.8ex] 
		& \multicolumn{2}{c}{Type~Ia} & \multicolumn{2}{c}{Type II} & \multicolumn{2}{c}{Type Ibc} & Total\\
		\hline\hline \\[-1.8ex] 
		$\mathcal{B}_\text{train}$ & 851 & (69.9\%) & 257 & (21.2\%) & 109 & (9.0\%)& 1,217 \\
		$\mathcal{U}_\text{train}$ & 292 & (24.3\%) & 749 & (62.4\%) & 159 & (13.2\%) & 1,200 \\
		$\mathcal{B}_\text{test}$ & 3,592 & (22.3\%) & 10,481 & (65.0\%) & 2,040 & (12.6\%)& 16,113 \\
		$\mathcal{U}_\text{test}$ & 4,151 & (25.7\%) & 9,989 & (61.9\%) & 1,990 & (12.3\%) & 16,130 \\
		\hline
	\end{tabular}
	\caption{Composition of training and testing datasets. \Btest\ is a realistic biased training set, while \Utrain\ is the `gold standard' unbiased training set. We compare the performance of our algorithm using both test sets to assess the effect of the biased training set on the classification quality.}
	\label{tab:datasets}
\end{table*}

\subsection{Modelling Light Curves with Gaussian Processes}\label{sec:modelling}

Let $X(t)$ be a stochastic process with continuous time index, $t$, in some time interval, $T$.
We say $X(t)$ follows a Gaussian Process (GP)  \citep[e.g.,][]{adler1990}, if the finite dimensional distribution, $p(X(t_1), \ldots, X(t_k))$, is (multivariate) Gaussian for any positive integer, $k$, and any set of time points $t_1, \ldots, t_k$ in $T$. Two key theoretical results are the existence and uniqueness of the Gaussian process. Specifically, a GP is uniquely determined by its mean function, 
\begin{equation}
\mu(t) =\E[X(t)]
\end{equation}
and its covariance function,
\begin{equation}
K(t,s) =\E \left[ \left\{ X(t) - \mu(t) \right\}^T \left\{ X(s) - \mu(s) \right\} \right],
\end{equation}
where $t$ and $s$ are any two time points in $T$. Conversely for any given mean and covariance functions,  there exists a GP with these mean and covariance functions. (For previous applications of GP regression to SN LC fitting, see ~\cite{Kim:2013bja}.)

The key result that allows us to use GPs to model time series such as LCs stems from the conditioning rule for multivariate Gaussian distributions \citep[e.g.,][]{rasmussen2006}. Suppose, for example, that $X$ follows 
a multivariate Gaussian distribution with mean vector $m$ and variance matrix $\Sigma$, i.e., $X \sim N (m, \Sigma)$, and partition
\begin{equation*}
X = 
\begin{Bmatrix}
X_1 \\
X_2
\end{Bmatrix},
\,
m = 
\begin{Bmatrix}
m_1 \\
m_2
\end{Bmatrix},
\text{ and }
\Sigma =
\begin{bmatrix}
\Sigma_{11} & \Sigma_{12} \\
\Sigma_{21} & \Sigma_{22}
\end{bmatrix}.
\end{equation*}
The conditional distribution of $X_2$ given $X_1$ is also a (multivariate) Gaussian, specifically  $X_2 \mid X_1 \sim N(m_*, \Sigma_*)$ with
\begin{equation} \label{e1}
\begin{aligned}
m_* &= \E[X_2|X_1] = m_2 + \Sigma_{21} \Sigma_{11}^{-1} (X_1 - m_1) \\
\Sigma_* &= \Var(X_2|X_1) = \Sigma_{22} -  \Sigma_{21} \Sigma_{11}^{-1} \Sigma_{12}.
\end{aligned}
\end{equation}

Turning to the modeling of LCs, let $f(t)$ denote an unobserved SN LC continuous in time. 
Suppose that 
\begin{equation}\label{eq:prior.GP}
f\sim GP(\mu, K),
\end{equation}
where $GP(\mu, K)$ denotes a GP with mean and covariance functions $\mu$ and $K$. (Here and elsewhere we suppress the dependence of $f$, $\mu$, and $K$ on time.)  In practice, we must specify the functional forms of $\mu$ and $K$, typically in terms of several unknown parameters. For the moment, we assume $\mu$ and $K$ are given, putting off discussion of their specification until Section~\ref{sec:MeanAndCov}. 

Because the distribution of $f(t)$ at any finite set of time points is multivariate Gaussian,  given a series of observations we can simply apply the formulas in (\ref{e1}) to obtain the conditional distribution of $f(t)$ at any other finite set of time points given the observed values. In this way, we can interpolate $f(t)$ between the observed values. Specifically, if we measure at $n$ points in time a vector of observations $\fobs = (f(t_1),\ldots, f(t_n))$, we can obtain the conditional distribution of $f(t)$ at another set of $k$ time points, namely  $\tilde f = (f(\tilde{t}_1),\ldots, f(\tilde{t}_k))$, by conditioning on the observations,
\begin{equation}\label{eq:post.GP}
\tilde{f} \mid \fobs =
\left.
\begin{pmatrix}
f(\tilde t_1) \\
\vdots \\
f(\tilde t_k)
\end{pmatrix}
\right|
\begin{pmatrix}
f(t_1) \\
\vdots \\
f(t_n)
\end{pmatrix}
\sim N_k \left(m_*, \Sigma_* \right),
\end{equation}
where $m_*$ and $\Sigma_*$ are in (\ref{e1}) with $m_1 = \left( \mu(t_1), \ldots, \mu(t_n) \right)^T$,  $m_2 = \left( \mu(\tilde{t}_1), \ldots, \mu(\tilde{t}_k) \right)^T$, $\Sigma_{11} = \mathbf{K}(t,t)$, $\Sigma_{12} = \mathbf{K}(t,\tilde{t})$, $\Sigma_{21} = \Sigma_{12}^T$, and $\Sigma_{22} = \mathbf{K}(\tilde{t},\tilde{t})$, where $\mathbf{K}(t,\tilde{t})$ is a matrix with element $(i,j)$ equal to $K(t_i,\tilde{t}_j)$. In Bayesian terms (\ref{eq:prior.GP}) is the prior distribution for $f$ and (\ref{eq:post.GP}) is the posterior distribution of $f(t)$ evaluated at a set of unobserved times.
Thus, we refer to (\ref{eq:prior.GP}) as the prior GP, to (\ref{eq:post.GP}) as the posterior GP, and to the vector $m_*$ as the posterior mean function. 
We use the posterior mean functions as the fitted LCs in our classification of SNe. 

Although the derivation of the posterior distribution in \eqref{eq:post.GP} does not account for measurement errors, they can easily be included in an elegant manner. This is a prime motivation for the use of GPs to model LCs. Assuming uncorrelated Gaussian errors, let $y_i$ and $\sigma_i$ denote the observed flux and standard error at time $t_i$, for $i= 1, \ldots, n$.  The data can be modeled as
\begin{equation} \label{e5}
y_i = f(t_i) + \sigma_{i} \varepsilon_i,
\end{equation}
where $f \sim GP(\mu, K)$ and $\varepsilon_i$ is iid $N(0,1)$ for $i=1,\ldots, n$. Thus, the posterior distribution of $f$ at a given set of $k$ time points $\tilde t_1,\ldots,\tilde t_k$ is simply
\begin{equation} \label{e3}
\tilde{f} | 
y \sim N_k (m_*, \Sigma_*),
\end{equation}
where $\tilde{f}= f(\tilde{t})$ and $m_*$ and $\Sigma_*$ are as in (\ref{eq:post.GP}) except that the noise vector $\sigma^2 = (\sigma^2_1, \ldots, \sigma^2_n)^T$ is added as a diagonal matrix to $\Sigma_{11}$ so $\Sigma_{11} = \mathbf{K}(t,t)+\text{diag}(\sigma^2)$. 

\subsection{Mean and Covariance Functions}\label{sec:MeanAndCov}
The choice of $\mu$ and $K$ can have a large influence on the quality of the LC interpolation in (\ref{eq:post.GP}).  The covariance function, for example,  controls the possible LC `shapes' that can be generated under the prior GP and hence depending on the amount and quality of data  may influence 
the posterior GP used for interpolation, namely the multivariate Gaussian distribution in (\ref{eq:post.GP}). 


In regions with abundant data the posterior GP is dominated by the observations, and the mean function has little influence. Therefore, the main focus of the literature is the covariance function, and the mean function is often simply taken to be a constant equal to zero, $\mu(t) = 0$. 
We adopt this choice and although the LC data often include sparse regions, we observe only limited problems with the posterior GP drifting towards the prior mean function, i.e., zero. 

There is a wide range of standard choices for covariance functions, each designed to model specific features of the data. These can be summed or multiplied \citep{rasmussen2006} to obtain a large and flexible strategy for specifying covariance structure \citep[see e.g.,][]{gelman2013}. 
Because data for the individual SN LCs are limited, such refined models are not appropriate. Instead we consider two relatively simple and flexible covariance functions, namely the squared exponential and the Gibbs kernels.

\subsubsection{Squared Exponential Kernel}
The squared exponential (SE) kernel is a popular choice \citep{roberts2012} and has been used before in fitting SN LCs \citep{Lochner2016,Kim:2013bja}. It is given by
\begin{equation} \label{eq:Kse}
K_{\rm se}(t,s) = \tau^2 \exp \left( - \frac{1}{2} \frac{(t-s)^2}{l^2} \right),
\end{equation}
where $\tau^2$ and $l$ are free parameters. The parameter $l$ is the length scale parameter and controls the speed with which the covariance decreases over time, i.e., how quickly the GP `forgets' previous observations and thus how rapidly it can fluctuate. The parameter $\tau^2$ is the variance parameter and controls the scale of the covariance function and hence the `amplitude' of the process. 
Notice that $K_{\rm se}(t,s)$ depends on $(t,s)$ only through the difference $|t-s|$, and hence corresponds to a stationary process. The kernel is continuous and infinitely  differentiable with respect to $r = t-s$. This means that the prior GP is smooth in the mean square sense, and we can expect the posterior GP and mean function to be smooth as well.

\subsubsection{Gibbs Kernel}\label{subsec:Gibbs}
The Gibbs kernel can be viewed as a generalization of the SE kernel in which the length scale parameter, $l$, is allowed to vary over time. 
This means that the timescale over which the GP remembers past observations and thus the timescale for variability can change. The added flexibility might be useful in accommodating different degrees of smoothing in different parts of the time domain, e.g., due to different levels of sparsity in the data. Specifically, the \citet{gibbs1997} kernel is given by,
\begin{equation}\label{eq:k.gibbs}
\begin{aligned}
& K_\text{Gibbs}(t,s)   = \\ 
& \tau^2 \left( \frac{2 l(t;\theta) l(s;\theta)}{l^2(t;\theta) + l^2(s;\theta)} \right)^{1/2}  \exp \left( - \frac{(t-s)^2}{l^2(t;\theta) + l^2(s;\theta)} \right),
\end{aligned}
\end{equation}
where $l(t;\theta)$ is a positive length scale function depending on time and a multivariate parameter $\theta$. We set 
\begin{equation}\label{eq:lengthscale}
l(t;\theta) = \lambda \left( 1 - p \, \varphi_{(t_\text{max}, \eta)}(t) \right),
\end{equation}
where $\lambda, \ p, \ t_\text{max},$ and  $\eta$ are tuning parameters and $\varphi_{(t_\text{max}, \eta)}(t)$ is a Gaussian density with mean $t_\text{max}$ and standard deviation $\eta$. With $p=0$ the Gibbs kernel reverts to the SE kernel.  For $t=s$, $K_\text{Gibbs} (t,t) = \tau^2$ and thus $\tau^2$ is a scaling parameter that controls the variance of the GP. 

\begin{figure}
	\centering
	\includegraphics[width=\linewidth]{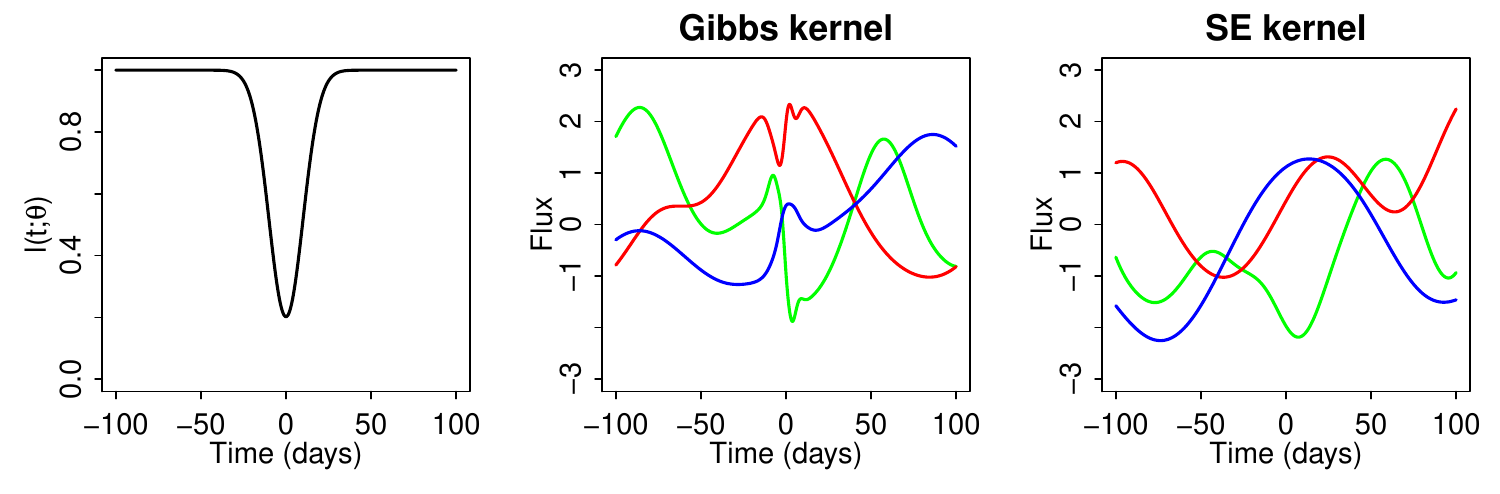}
	\caption{\emph{Left}: The length scale function for the Gibbs kernel with $\lambda=1$, $p=20$, $t_\text{max}=0$ and $\eta = 10$. \emph{Middle}: Three functions drawn from the GP prior, $\GP(0,K_\text{Gibbs})$, with the Gibbs kernel and length scale function in the left plot and with $\lambda = 20$ (green), $\lambda=30$ (red), and $\lambda =40$ (blue). \emph{Right}: Three functions drawn from $\GP(0,K_{\rm se})$ with $\tau=1$ and $l=20$ (green), $l=30$ (red), and $l=40$ (blue).} 
	\label{fig33} 
\end{figure}

An example of $l(t;\theta)$ with $\lambda=1$, $p=20$, $t_\text{max}=0$ and $\eta = 10$ is plotted in the left panel of Figure~\ref{fig33}. Three functions drawn from $\GP(0,K_\text{Gibbs})$ are plotted in the middle panel (using the $l(t;\theta)$ as plotted in the left panel) and three drawn from $\GP(0,K_\text{se})$ are plotted in the right panel of Figure~\ref{fig33}. Clearly, the Gibbs kernel allows more rapid fluctuations around $t=0$. The functions plotted in the two right panels of Figure~\ref{fig33} can be viewed as draws form a prior GP. As always, with sufficient data the influence of the choice of kernel on the posterior GP is expect to vanish.  A further comparison of the GP LC models fitted with the two kernels appears in Section \ref{subsec:gibbs.fitting}.

\subsection{Fitting the Gaussian Processes}\label{subsec:max}

The GPs defined by the kernels in \eqref{eq:Kse} and \eqref{eq:k.gibbs} have free parameters that need to be fit from the LC data. A natural way to handle the parameters of the covariance functions is to assign them hyperprior distributions. Letting $\theta$ denote the collection of parameters of the kernel, the posterior distribution of $\theta$ is
\begin{equation} \label{e2}
p(\theta | y, t) \propto p(y | \theta, t) p(\theta | t),
\end{equation}
where $p(\theta | t)$ is the prior distribution for $\theta$ and $p(y | \theta, t)$ is the marginal likelihood,
\begin{equation} \label{e4}
p(y|t, \theta) = \int p(y,f|t, \theta) {\rm d}f = \int p(y|f, t, \theta) p(f| t, \theta) {\rm d}f,
\end{equation}
where the integration marginalizes out the dependence of $y$ on the latent functions, $f$. We refer to the distribution in (\ref{e2}) as simply `the posterior', which should not to be confused with the ``posterior GP'' given in  (\ref{e3}). 

\subsubsection{MAP Estimation}

We aim to use the posterior in (\ref{e2}) to estimate the mean function of the LCs in order to classify the SNe. For this purpose a single reasonable estimator of the mean function is sufficient, that is, it is not necessary to carefully quantify uncertainty in the estimate. We use a strategy common in machine learning that uses the maximum a posteriori probability (MAP) estimate, which is the value of $\theta$ that maximizes the posterior distribution in (\ref{e2}).

To compute the MAP estimate of $\theta$, we first need to derive an expression for the marginal likelihood in (\ref{e4}). This can be accomplished analytically, because the integrand in (\ref{e4}) can be written as the product of two Gaussian densities, specifically, $y|f, t, \theta \sim N(f, \text{diag}(\sigma^2))$ and, given the discrete set of time points, $t$, we have, $f| t, \theta \sim N\{\mu(t),$ $\mathbf{K}(t,t) \}$.
Letting $t= (t_1,\ldots, t_n)^T$, $K=\mathbf{K}(t,t)$, $\mu = \mu(t)$ and $\Sigma = \text{diag}(\sigma^2)$, we thus obtain the marginal likelihood,
%
%
\begin{equation}\label{e6}
\begin{aligned}
\log p(y|t, \theta) & = -\frac{1}{2} (y-\mu)^T (K+\Sigma)^{-1}(y-\mu)  \\
- & \frac{1}{2} \log |K+\Sigma| - \frac{n}{2} \log{2\pi}. 
\end{aligned}
\end{equation}
Only the first term in (\ref{e6}) involves the data. The second is a penalty for the complexity of the fit. Under the SE kernel, for example, a less complex approximately linear fit has a higher value of $l$ causing the exponential term in (\ref{eq:Kse}) to be nearly constant regardless of the distance $(t-s)$. In this case the determinant in the second term of (\ref{e6}) is dominated by the diagonal matrix, $\Sigma$, because $K$ is nearly linearly dependent. The final term in (\ref{e6}) is a normalizing constant that depends on the data only through $n$.

To obtain the log of the posterior in (\ref{e2}) the log-prior can simply be added to (\ref{e6}). Because (\ref{e6}) only depends on $\theta$ through $K$, its derivatives are easily obtained, numerical optimization of (\ref{e2}) is efficient, and it can be easily implemented using gradient-based numerical methods.\footnote{We use the constrained gradient based optimizer developed by \citet{byrd1995} and implemented in \texttt{R} through the \texttt{optim()} function. This optimizer is based on the quasi-Newton method and only requires the Jacobian (or in our case the gradient), but not the Hessian to be calculated. Instead the Hessian is approximated \citep[sec. 2.2.2.3]{givens2012}.}

\begin{figure}
	\centering
	\includegraphics[width=\linewidth]{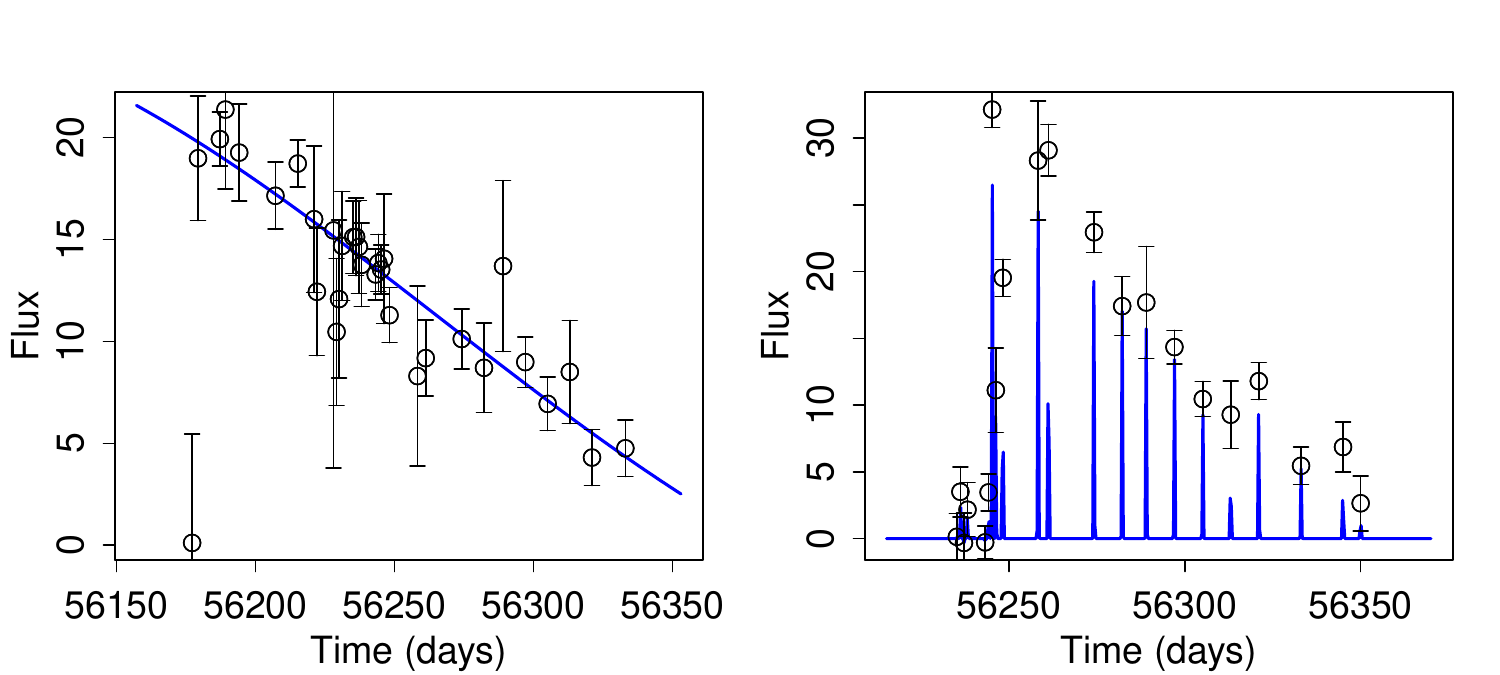}
	\caption{Examples of fitted LCs (i.e. the posterior mean function in (\ref{e3})) showing pathological MAP estimates under the SE kernel and a flat prior on the kernel's parameters. The fitted LCs are plotted in blue. The left panel is an examples of a very large fitted value of $l$ ($r$-band of SN~84341 at $z=0.95$, with $l=247.2$ and, $\sigma=18.4$). The right panel is an example of a very small fitted value of $l$ ($r$-band of SN~92200 at $z=0.41$, $l=0.1$, and $\sigma=15.1$.}
	\label{fig13} 
\end{figure}

\subsubsection{Prior Distributions and Fits Under the SE Kernel} \label{subsubsec:fitting}

\citet{Lochner2016} fitted LCs using a flat prior distribution on the parameters of the covariance function in (\ref{e2}) \citep{seikel2012}. This is a common choice in the GP literature (e.g.~\citet{rasmussen2006}, sec.~5.4.1). Although there is in no general guarantee that the posterior is proper with an improper prior, this causes us no difficulty in principle because we only use the MAP estimate and the marginal likelihood is finite. 

\begin{figure}
	\centering
	\includegraphics[width=\linewidth]{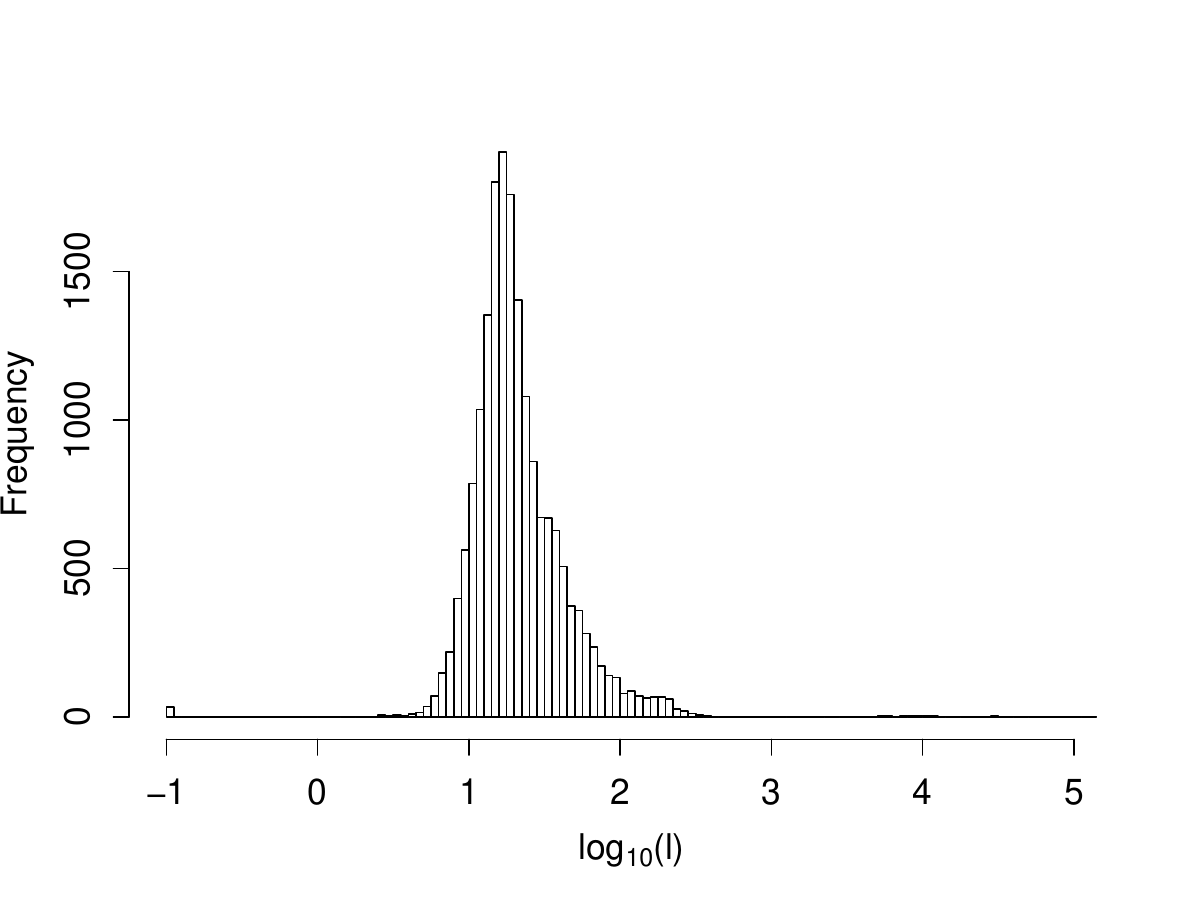}
	\caption{Distribution of the fitted $l$ parameters (MAP estimates) of the $r$-band fitted with the uniform (improper) prior on $l$. Notice the extreme values on the log-scale.}
	\label{fig11}
\end{figure}

Unfortunately, using a  flat prior gives rise to pathological fits with the data we used. Two examples are plotted in Figure~\ref{fig13}. The left plot is an example of a large fitted value for $l$, causing an almost linear fit. The right plot is an extreme example of problems that occur when the posterior process drifts towards the prior mean, in this case caused by a very small fitted value of $l$. 
The histogram in Figure~\ref{fig11} gives the distribution of the fitted (i.e., MAP) values of $l$ in the $r$-band across the 17,730 SNe (on the log-scale). Extreme values of $l$ in this distribution lead to poor fits like those illustrated in Figure~\ref{fig13}.
The other bands suffer from similar problems. 

The flat prior distribution on $\tau$, on the other hand, does not cause problems. To see this, we rescaled the data by dividing each LC by its largest value, adjusted each $\sigma_i$ accordingly, and plotted GPs with extreme values in the distribution of   the fitted values of $\tau$. No systematic patterns were observed.

To avoid pathological cases like those in Figure~\ref{fig13} requires a proper prior distribution on $l$.  The empirical distribution in Figure~\ref{fig11} obtained using a flat prior inspired the adoption of a log-normal prior distribution,
\begin{equation} \label{e10}
p(l| \nu, \rho) =\frac{1}{\rho {\sqrt {2\pi }}}  \frac {1}{l} \exp\left(-{\frac {\left(\log l-\nu \right)^{2}}{2\rho ^{2}}}\right).
\end{equation}
The distribution has support on the positive real line is right skewed and has, depending on the scale parameter, $\nu$, very low probability for small values. We set the hyper parameters to $\nu=3.1$ and $\rho=0.4$, yielding a mean of $\approx 24$ and median of $\approx 22$, close to the mean and median of the fitted parameters in all bands obtained using a flat prior on $l$. Using this prior avoid the pathological fits without interfering too much with the GP fits where the flat prior works well. The flat prior on $\tau$ is maintained. 

\begin{figure*}
	\centering
	\includegraphics[width=\linewidth]{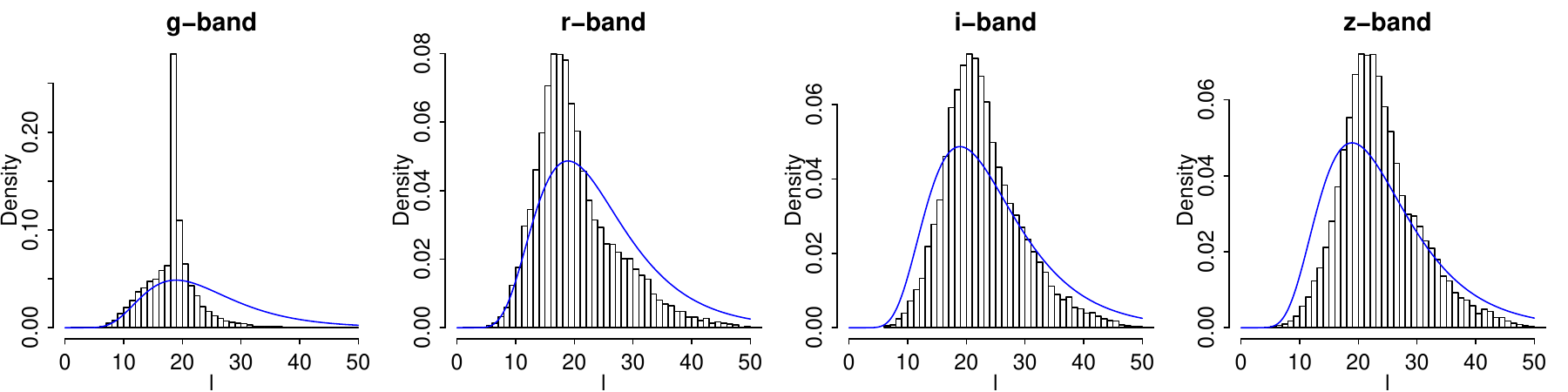}
	\caption{Histograms of fitted $l$ parameters of the SE kernel using the log-normal prior on $l$, for the entire data set. Superimposed in blue is the log-normal prior distribution.}
	\label{fig14} 
\end{figure*}

The distributions of the fitted values of $l$ across the 17,730 SNe under the log-normal prior distribution are given in Figure~\ref{fig14}, for each of the four bands. These histograms are not plotted on the log scale as in Figure \ref{fig11} and illustrate that this choice of prior avoids the extreme fitted values of $l$ and their associated pathological fitted LCs. The distributions of the MAP estimates of $l$ are quite similar to the prior distribution superimposed on the histograms. This is not a histogram of the posterior distribution of $l$ for a single LC, but the distribution of the $l$ across the LCs. Thus, the similarity between the histograms and the priors does not indicate an overly influential prior distribuiton, but rather that the prior distribution is a relatively good representation of the population distribution of $l$ across SNe (as we would expect, for example, in a hierarchical model). The one exception is the $g$-band, where the peak in the histogram near the prior mode can be explained by the fact that at higher redshift the $g$-band maps deeper into the UV, where the SNIa flux drops, leading to fainter emission. This produces more noisy data in this band for many SNe, similar to the SN plotted in Figure \ref{fig1}. With higher noise, the prior dominates the likelihood and hence the posterior distribution peaks near the prior mode. 

\subsubsection{Prior Distributions and Fits Under Gibbs Kernel}\label{subsec:gibbs.fitting}
The Gibbs kernel and its length scale function given in (\ref{eq:k.gibbs}) -- (\ref{eq:lengthscale}) have five free parameters. Initial investigations showed that with the Gibbs kernel the log marginal likelihood in (\ref{e6}) often contains many local maxima. To regularize the fits, we restrict the parameter values, either deterministically or through prior distributions. Some local modes, for example, correspond to the the position parameter, $t_\text{max}$, being in regions with noisy data that are away from the peak brightness. This results in over fitting of the noisy regions of the LCs. To avoid this, we restrict $t_\text{max}$ to be equal to the value of $t$ that maximizes the fitted LCs under the SE kernel. This restriction does not imply that the Gibbs and SE kernel fits necessarily exhibit a maximum at the same time value, but such values are likely to be similar. We also fix the variance parameter at $\eta=10$ as in the leftmost panel of Figure \ref{fig33}, restrict $p$ to the interval $[0,20]$, and restrict $\lambda$ and $\tau$ to be positive. The restriction on $p$ ensures that length scale function, $l(t; \theta)$, is positive.

Since the SE kernel is a special case of the Gibbs kernel (with $p=0$), the considerations regarding the choice of prior for $l$ discussed in Section~\ref{subsubsec:fitting} also apply to the scale length, $\lambda$, under the Gibbs kernel. 
Hence we adopt the same log normal prior distribution used for $l$ under the SE kernel for $\lambda$. Finally, we use uniform prior distributions for $p$ and $\tau^2$. Because numerical optimizers can easily converge to one of the local maxima, for a number of LCs we obtain a lower value for the log posterior MAP estimate than we did with the SE kernel. In these cases, the SE fits are used instead. (I.e., we set $p=0$, $\lambda=l$, and $\tau_\text{Gibbs}=\tau_{\rm se}$.) The log posterior of the MAP estimates increases when using the Gibbs kernel (as compared with the SE kernel) for 68\%, 75\%, 57\% and 51\% of the LCs in the $g$, $r$, $i$ and $z$ bands, respectively. 

\subsubsection{Comparing the Fits}

\begin{figure}
	\centering
	\includegraphics[width=\linewidth]{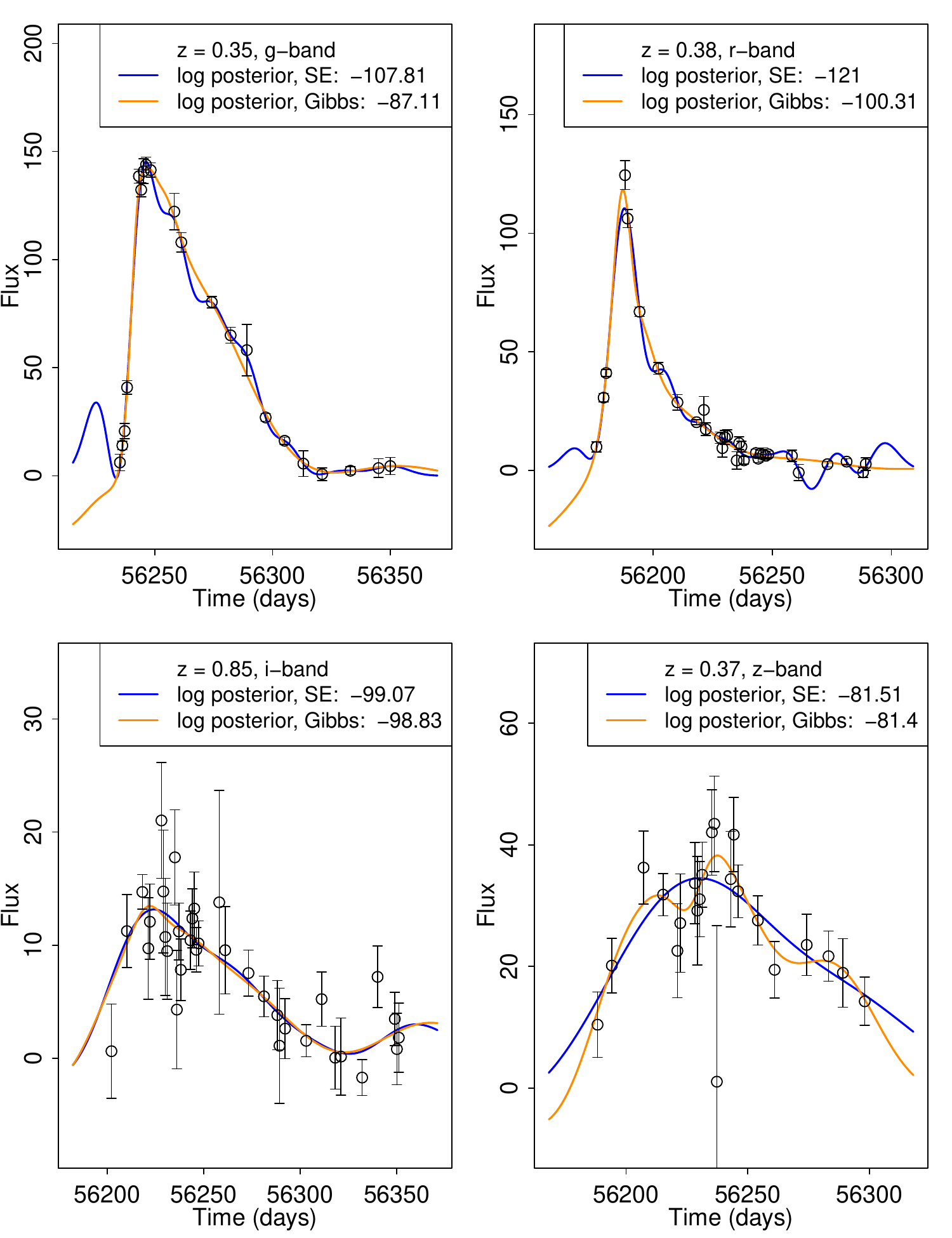}
	\caption{\emph{Top row}: Two examples of LCs ($g$-band of SN~17125 and $r$-band of SN~78789, respectively) for which the Gibbs kernel (orange) provides a better fit than does the SE kernel (blue). In both cases the Gibbs kernel is able to fit the peak well without over fitting away from the peak, whereas the SE kernel results in a small length scale and over fits. In both cases the log posterior improved substantially with the Gibbs kernel. \emph{Lower left}: A LC ($i$-band of SN~113107) where the two fits are quite similar. As expected, the log posterior is similar under the two kernels. \emph{Lower right}: A LC ($z$-band of SN~147174) where the log posterior of the two fits are almost identical, but the fit under the Gibbs kernel arguably overfits the noisy data. \emph{Fitted Parameters}: Upper left ($g$-band of SN~17125), SE fit with $l=7.24$ and $\tau = 58.76$, Gibbs fit with $\lambda = 35.21, \tau = 75.65$, and $p=20$; Upper right ($r$-band of SN~78789), SE fit with $l=7.6$ and $\tau=34.75$, Gibbs fit with $\lambda=36.46, \tau=50.48$, and $p=20$; Lower left ($i$-band of SN~113107), SE fit with $l=23.63$ and $\tau = 6.23$, Gibbs fit with $\lambda = 30.91, \tau = 6.5$, and $p=9.66$; Lower right ($z$-band of SN~147174), SE fit with $l=34.35$ and $\tau=17.64$, Gibbs fit with $\lambda=22.11, \tau=20.62$, and $p=8.73$. 
    }
	\label{fig16} 
\end{figure}

Figure~\ref{fig16} depicts four representative LCs and their fits under both kernels. The two panels in the first row are good examples of cases that motivate the Gibbs kernel. In both cases, the SE kernel is unable to simultaneously fit the rapid increase followed by a rapid decrease in brightness at the peak and the relatively smooth tails. The result is a relatively small fitted value of $l$ and thus too little smoothing in the tails. The Gibbs kernel in contrast benefits from the flexible length scale function and is able to fit both the peak and the tails well. The LC at the bottom left is an example where both kernels yield very similar fits (although the fitted value of $p$ is not near 0 under the Gibbs kernel). The bottom right LC is an example where the increased flexibility of the Gibbs kernel results in overfitting of relatively noisy data. 

\begin{figure*}
	\centering
	\includegraphics[width=\linewidth]{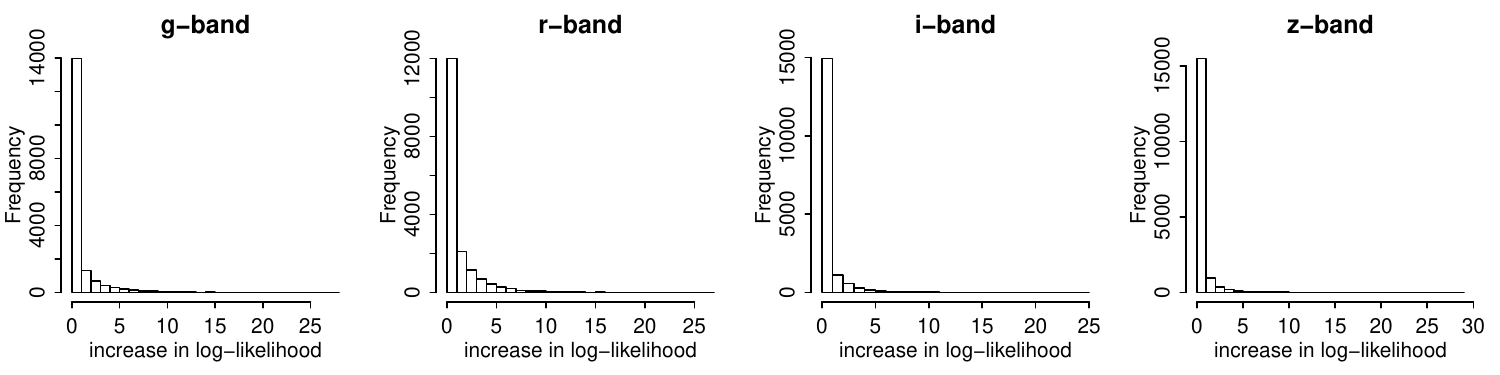}
	\caption{Histograms of the difference between the maximum log posterior obtained using the Gibbs kernel and that obtained using the SE kernel. Values greater than zero indicate an improved fit under the Gibbs kernel. (In cases where the log posterior obtained with the Gibbs kernel is less than that obtained with the SE kernel, a zero is recorded in these histograms.)
    }
	\label{fig17} 
\end{figure*}

Histograms of the difference between the maximum log posterior obtained using the Gibbs kernel  and that obtained under the SE kernel appear in Figure~\ref{fig17}.
(In cases where the log posterior obtained with the Gibbs kernel is less than that obtained with the SE kernel, a zero is recorded in these histograms.) 
In most cases, the increase in the maximum log posterior value achieved under the Gibbs kernel is small (and thus may be included in the leftmost  bins  of the histograms in Figure~\ref{fig17}.)

Cases of overfitting under the Gibbs kernel, like the one depicted in the bottom right panel of Figure~\ref{fig16}, are typically associated with noisy data. In such cases the fits under the two kernels appear quite different while their maximum log posterior values are quite similar. 
\edit{In principle a formal model selection technique (e.g., a Bayes Factor or likelihood ratio test) could be employed to select between the SE and Gibbs kernel but this would require substantial computational effort. Since we do not believe either kernel is ``correct'' and simply aim for reasonable interpolation of the LCs, we propose a simple method to choose between the kernels. Specifically, to avoid the occasional overfitting associated with the Gibbs kernel, we use this kernel only if it} improves the value of the log posterior MAP by at least 2\% and using the SE kernal otherwise. In this way we aim to use the Gibbs kernel when it is advantageous (e.g., the cases in the first row of Figure~\ref{fig16}), but not when it overfits (e.g, the case in the lower right panel of Figure~\ref{fig16}).  This strategy results in the Gibbs kernel being used for 
 17\%, 28\%, 12\% and 8\% of the LCs, in the $g$, $r$, $i$ and $z$ bands, respectively. In our numerical studies in Section~\ref{sec:results} we compare this strategy with (i) using the SE kernel for all SNe and (ii) using the Gibbs kernel for all SNe.

\section{Normalising the Light Curves}\label{sec:norm}
The fitted LCs must be aligned in time and normalized in brightness before they are classified.  We describe both of these tasks in this section.

\subsection{Time Alignment}\label{sec:defining.t0}
To align the LCs in time we define time zero to be the time when the fitted LC in the (observer's) $i$-band achieves its maximum.\footnote{This differs from the more standard practice of defining time zero as the time of peak brightness in the rest-frame $b$-band. Because the $b$-band is not available in the current data, converting the LCs to the SN rest-frame would require complex K-corrections and accounting for uncertainty in the redshift. Thus, using the observer's frame and the $i$-band significantly simplifies our procedure.}
We choose the $i$-band because it is the band where most of the fitted LCs (under the SE kernel) have a maximum \emph{within} the time interval of the observations, namely 90.4\% of the SNe LCs peaked in the $i$-band.

\subsubsection{Estimating Time Zero for LCs with a Peak}\label{subsec:Estimating.T0.with.a.peak}
We evaluate the fitted LCs on a one-day spaced grid of time points, starting at the time of the first observation in the $i$-band. Given that the fitted LCs do not fluctuate much on a daily basis and the computational costs involved with a finer grid, we believe this choice of grid resolution offers a reasonable compromise between computational cost and accuracy. Let $\hat{f}_a (t_j)$ denote the fitted LC of SN $a$ at time $t_j$ in the $i$-band. (The $i$-band is suppressed in this notation.) Letting $t_{a,f}$ and $t_{a,l}$ be the times of the first and last observations of SN $a$ in the $i$-band, respectively, and $\mathcal{T}_a = \{ k \in \mathbb{N} \, | \, 0 < k \leq t_{a,l} - t_{a,f} \}$, we define time zero as 
\begin{equation}\label{eq:t0}
t_{a,0} = t_{a,f} + \argmax_{ k \in \mathcal{T}_a } \hat{f}_a (t_{a,f}+k).
\end{equation}

\subsubsection{Estimating Time Zero for LCs without a Peak}\label{subsec:estimating.t0}

The LCs of some of the data were generated with explosion times well before or well after the survey time window, leading to several LCs with missing pre- or post-peak data.
Of our 17,330 SNe, after evaluating the fitted LCs on one-day spaced grids (Section \ref{subsec:Estimating.T0.with.a.peak}) 1,721
do not have a peak in the $i$-band. For these, time zero is estimated via pairwise comparisons with the peaked LCs. Consider, for example, a pair of SNe, where one has a peak in the $i$-band and one does not.
The LC without a peak is repeatedly shifted according to the one-day spaced grid and its time zero
is estimated based on the shift that minimises the mean squared distance between the two LCs. For each LC without a peak, an estimate of this sort is obtained using each of the 15,609 LCs with a peak and the estimates are combined using a weighted average to obtain the final estimate.

We start by considering LCs for which we have data recorded only after the peak. Let $a$ be the index of one such SN and let $b$ be the index of one of the SN with a peak in the $i$-band. To compare the two LCs, we normalize SN $a$ to its fitted value at the time of its first observation,
$\hat{f}_a(t_{a,f})$, and normalize SN $b$ to its fitted value $k$ days past its peak, $\hat{f}_b (t_{b,0}+k)$. (Recall that we aim to estimate the number of days between the unobserved peak for SN $a$ and its first observation.) We then compute the mean square error of the rescaled LCs with a time shift of $k$ days,
\begin{equation}\label{eq:cij}
c_{ab}(k) := \frac{1}{n_a} \sum_{r = 0}^{n_a-1} \left( \frac{\hat{f}_a(t_{a,f}+r)} {\hat{f}_a(t_{a,f})} - \frac{\hat{f}_b(t_{b,0} + k + r)}{\hat{f}_b (t_{b,0}+k) } \right)^2,
\end{equation}
where $n_a = \left \lfloor t_{a,l} - t_{a,f} \right \rfloor$ is the number of grid points for SN $a$. The estimate of $t_{a,0}$ based on the LC of SN $b$ is the shift (in days) that minimizes the mean square error, 
\begin{equation}\label{eq:estij}
\hat{t}^\text{shift}_{a,b} = \argmin_{k \in \{0,1,\ldots, n_{ab} \} } c_{ab}(k), 
\end{equation}
where $n_{ab} = n_b - n_a$ is the difference between the number of grid points. The range of values of $k$ in (\ref{eq:estij}) ensures that $\hat f_a$ is evaluated only where it exists in (\ref{eq:cij}).  If $n_{ab} < 0$, $\hat{t}^\text{shift}_{a,b}$ is set to 0. 

Our estimate of time zero for SN $a$ is a weighted average of $\hat{t}^\text{shift}_{a,b}$ across all $b$ for which SN $b$ is observed with peak in the $i$-band; let $\mathcal{P}$ be the collection of indexes, $b$, of such SNe. In order to favour peaked SNe that (after a shift) align better with SN $a$, the weights are chosen to be inversely proportional to the mean square error 
\begin{equation}\label{eq:min}
c_{ab} =  \left\{\min_{k \in \{0,1,\ldots, n_{ab} \} } c_{ab}(k)\right\}^{-1},
\end{equation}
and our final estimate of time zero is
\begin{equation}\label{eq:ave}
{t}_{a,0} = t_{a,f} - \frac{1}{\sum_{b \in \mathcal{P}} c_{ab}} \sum_{b \in \mathcal{P}} c_{ab} \, \hat{t}^\text{shift}_{a,b}.
\end{equation}
(Again, if $n_{ab} < 0$, $c_{ab}$ is set to 0.)

A similar method is used to estimate time zero for the LCs observed only before their peaks in the $i$-band. The only difference is that the LCs in (\ref{eq:cij}) are scaled to have a common right end point, the summation in (\ref{eq:cij}) is from $r=-n_a+1$ to $0$ and the minima in (\ref{eq:estij}) and (\ref{eq:min}) are taken over $\{-n_{ab},-n_{ab}+1, \ldots, 0 \}$. This situation happens only for  5.4\% of the SNe. 

Our method is inspired by and extends that adopted by \citet{Richards2012}, who used the shift that minimised the cross correlation between the two LCs rather than (\ref{eq:cij}) and (\ref{eq:estij}), and adopted a simple average rather than a weighted average. 

\subsubsection{Standardising the Time Domains}
We use our estimate for time zero to re-center the time scale for each SN (in all colour bands) at the point of its maximum $i$-band brightness. Specifically, for each SN, with index $a$, define a standardized, integer sequence of times in units of days, $(\tilde{t}^\band_{a,1}, \ldots, \tilde{t}^\band_{a, n_a^\band})$, where
\begin{equation*}
\tilde{t}^\band_{a,1} = \left \lceil t^\band_{a,f} - t_{a,0} \right \rceil \ \hbox{ and } \
\tilde{t}^\band_{a, n_a^\band} = \left \lfloor t^\band_{a,l} - t_{a,0} \right \rfloor  
\end{equation*}
where  $n_a^\band = \tilde{t}^\band_{a, n_a^\band} - \tilde{t}^\band_{a,1} + 1$ and the superscript, $\band$, indexes the colour bands, $\band \in \bandset = \{g,r,i,z\}$. 

\begin{figure}
	\centering
	\includegraphics[width=\linewidth]{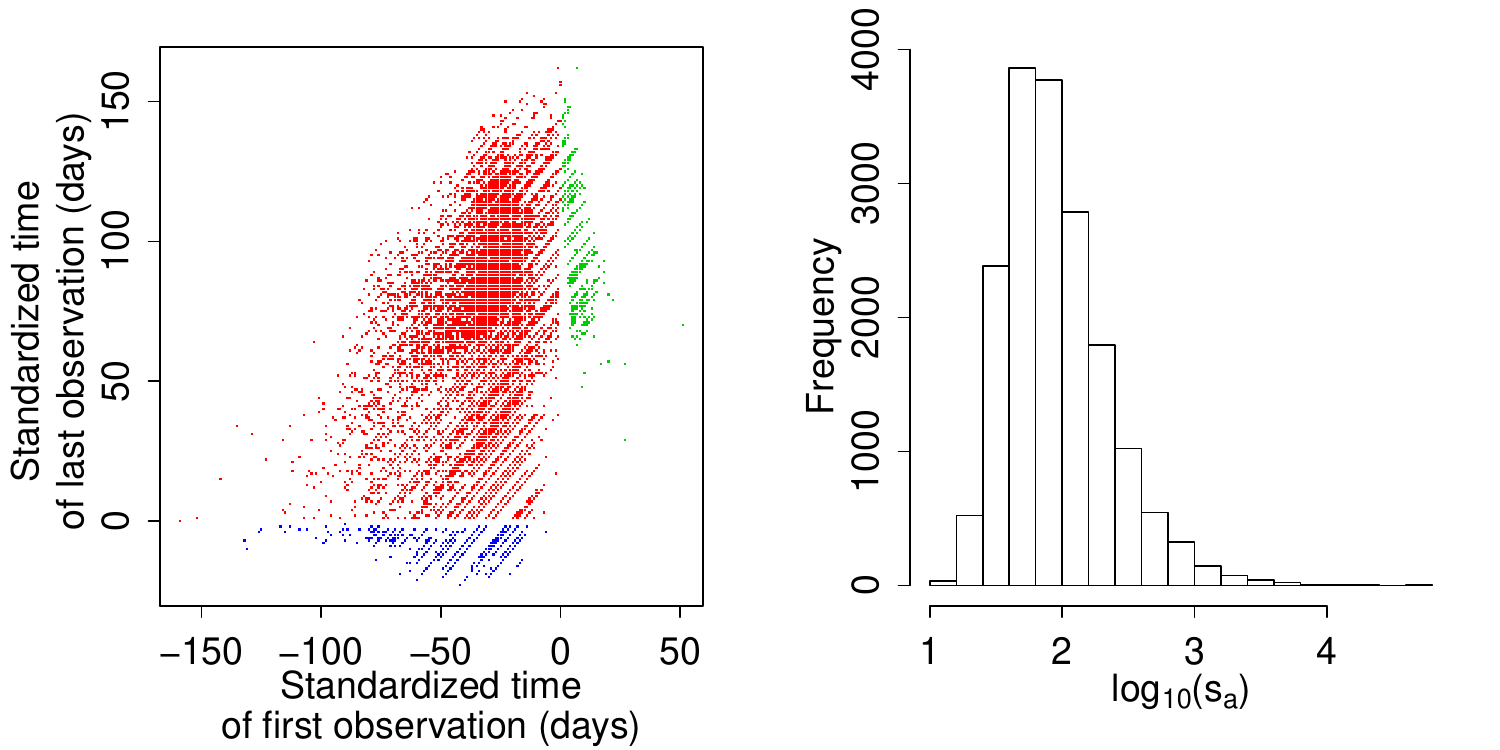}
	\caption{\emph{Left:} Scatterplot illustrating the observed time intervals of the $i$-band LCs in standardised times, $\tilde{t}$. The plot is colour coded according to whether the $i$-band peak occurs within the observed LC: blue indicates data is only recorded before the peak, green indicates that data is recorded only after the peak, and red indicates that the peak is observed. \emph{Right:} Histogram of the SNe apparent brightnesses, $s_a$ (arbitrary units), on log-scale (calibrated fluxes are obtained by converting from apparent magnitudes using Eq.~(1) in  \citet{kessler2010}). Both plots are based on the SE kernel.}
	\label{fig20} 
\end{figure}

In left panel of Figure~\ref{fig20} we plot the standardized times of the first and last observation in the $i$-band to illustrate the time intervals of the observed LCs. The plot is obtained using the SE kernel.


\subsection{Standardising the Fluxes}\label{sec:standardising.fluxes}
Renormalizing the LCs onto a common brightness scale accentuates certain features that aide classification. (We identify these features using diffusion maps as discussed in Section~\ref{sec:diff.maps}.) Following \citet{Richards2012}, we divide the LCs of each SN by the sum of the maxima in all of its bands.  Specifically, the normalised LC of SN $a$ at time $t_j$ in band $\band$ is
\begin{equation}\label{eq:normalising.flux}
\tilde{f}^\band_a(t_j) = \frac{\hat{f}^\band_a (t_j)}{s_a}, 
\ \hbox{ with } \
s_a = \sum_{\band \in \bandset =\{g,r,i,z\}} 
\max_{k} \hat{f}^\band_a (\tilde{t}^\band_{a,k} + t_{a,0}),
\end{equation}
where the maximum is over 
$k \in \left\{\tilde{t}^\band_{a,1}, \ldots, \tilde{t}^\band_{a, n_a^c}\right\}$. 
{Thus $s_a$ is the sum of the maxima of the LCs in the four bands. Since the maxima occur at slightly different times for different bands, this quantity does not exactly represent the peak apparent brightness, but acts as a rough proxy for it.} A log-scale histogram of the $s_a$ is shown in the right panel of Figure~\ref{fig20} and illustrates the distribution of overall apparent brightnesses among the SNe. (We refer to the $s_a$ as the ``brightnesses''.) The histogram is based on the fit of the SE kernel.

We denote the set of normalised LCs for SN $a$ by
\begin{equation}\label{eq:Yi}
Y_a = \left\{ \left(\tilde{t}^\band_{a, j}, \tilde{f}^\band_{a,j}\right) \left|  \band \in \{g,r,i,z\}, j=1,\ldots, n_a^\band \right. \right\},
\end{equation}
where $\tilde{f}^\band_{a,j} = \tilde{f}^\band_a(\tilde{t}^\band_{a, j}+ t_{a,0}))$.

%

\section{Classification Methodology}\label{sec:classification}



\subsection{Diffusion Maps}\label{sec:diff.maps}

The observed SNe are divided into a training and a test set, where the SN types are known in the training set and we aim to identify the types in the test set by comparing SN LCs. Although our GP fits to the LCs allow extrapolation to a common time interval, the degree of variability among the observation intervals (see left panel of Figure~\ref{fig20}) means that expanding to a common interval would require substantial extrapolation. 

To avoid this, we follow \citet{Richards2012} and apply a diffusion map. This allows us to compute a vector of length $m$ for each LC in such a way that the Euclidean distance between these vectors is a measure of similarity between the LCs. (Formally the distance approximates the so-called diffusion distances between two LCs, as defined in Section~\ref{ssec:diff.maps}.) This allows classification of the SNe to be accomplished by applying any ``off-the-shelf'' classification algorithm to the $m$ dimensional vectors, which thus act as predictor variables for the classification. As discussed in Section~\ref{sec:random.forest}, we use a random forest classifier. 

The key advantage of this strategy is that it only requires {\it pairwise} comparison of the LCs. Thus, the time intervals on which the LCs are compared can be different for each pair; we choose the intersection of the observations intervals for each pair. Although the intersection may be short (or even empty), a further advantage is that the diffusion distance between two LCs is based on the average difference in distances between the two and all other LCs. That is, the diffusion distance between LC $a$ and LC $b$ is a function of the difference in distances $a\leftrightarrow h$ and $b \leftrightarrow h$ averaged over all LCs $h$. Thus, as long as the observation intervals are not completely separable over the entire dataset, each pair of LCs can be compared in the diffusion space.  

In contrast to \citet{Richards2012}, we apply the diffusion map to each colour band separately. Thus, for each SN, we obtain four vectors of predictor variables (one for each band). The lengths of these vectors may vary among the bands, and thus we denote the lengths by $m_\band$, for $\band \in \bandset = \{g,r,i,z\}$. The advantage of this approach is that the explanatory power of the LCs in the different bands are not blurred with each other. Thus two similar SNe with one or more bands disrupted by noise can still be judged similar by using the unaffected bands. An overview of diffusion maps and the details of our choice of the diffusion distance are given in Appendix~\ref{app:diff.maps}.

\subsection{Random Forest Classification}\label{sec:random.forest}

Random forest is a classification algorithm that is based on ``growing'' replicate classification trees using bootstrapped samples of the data. Each classification tree is grown by recursively splitting the data into binary regions based on one of the predictor variables. (Recall the predictor variables are generated by the diffusion map in our case.) In each split, the variable and split point is chosen such that the sum of the \emph{gini} indexes\footnote{With two classes, the gini index of a region is defined as\break
$2p_\text{Ia} (1 - p_\text{Ia})$, where
$p_\text{Ia}$ is fraction of SN that are SNIa in the region. Thus, the gini index will be minimised if nearly all or nearly none of the SNe in the region are SNIa.} in the two new regions is decreased as much as possible. In each split only a randomly selected subset of $m_\text{try}$ of the predictor variables are considered.  The trees are grown to their maximum size, which means that the recursive partitioning of the predictor variables into sub-regions is stopped only when one class (of SNe) is present in each region, e.g. because there is only one SN left in the region. Once all of the trees are grown, the classification of SN from the test set is based on `voting' of the trees. Since we have only two classes, SNIa and not SNIa, we can use a simple threshold, $\gamma_\text{Ia}$, where we classify a SN as SNIa if the proportion of trees so voting is greater than $\gamma_\text{Ia}$.



Random forests are relatively robust to noisy predictor variables as long as there is a reasonable number of relevant variables \citep{friedman2009}. Therefore, all 
the predictor variables from the four optimised diffusion maps (i.e., up to 100) are combined into a vector and passed to the random forest without further selection. 

\subsection{Tuning the Diffusion Maps and Random Forest Classifier}\label{sec:tuning}

To implement the diffusion map and random forest, we must set their tuning parameters, $(\varepsilon_g, \varepsilon_r, \varepsilon_i, \varepsilon_z)$, $m_\text{try}$, and $\gamma_\text{Ia}$. (See (\ref{eq:w.func}) in Appendix~\ref{app:diff.maps} for a definition of $\varepsilon_\band$; there is a separate value of $\varepsilon_\band$ for each colour band,  $\band\in\bandset$.) We aim to set these parameters in such a way as to optimise classification, while recognizing that classifying a SN as a SNIa when it is not (i.e. a false positive) is worse than neglecting to identify a SN as a SNIa (i.e., a false negative), at least in the typical application of using SNIa as standard candles. For this reason, \citet{kessler2010} proposed the following criterion:
\begin{equation}\label{eq:zeta.Ia}
\zeta_\text{Ia} (W) = \frac{N_\text{Ia}^\text{True}}{N_\text{Ia}^\text{Total}} \times \frac{ N_\text{Ia}^\text{True} }{N_\text{Ia}^\text{True} + W N_\text{Ia}^\text{False}},
\end{equation}   
where $N_\text{Ia}^\text{Total}$ is the total number of SNIa, $N_\text{Ia}^\text{True}$ is the number of correctly classified SNIa (true positives), and $N_\text{Ia}^\text{False}$ is the number of SNe incorrectly classified as a SNIa (false positives). The first term on the right hand side of (\ref{eq:zeta.Ia}) is also known as the efficiency, $e_\text{Ia}$, of the classifier, i.e., the proportion of the SNIa that are correctly identified. For $W=1$, the second term of (\ref{eq:zeta.Ia}) is known as the purity, $p_\text{Ia}$, i.e., the proportion of true positives among the overall number of positives. In the classification challenge $W$ was set to 3 \citep[Section~5]{kessler2010} meaning that false positives incur a heavier penalty than false negatives. For simplicity, we fix $\zeta_\text{Ia}=\zeta_\text{Ia} (3)$.

We set the tuning parameters in two steps. First we set each of the $(\varepsilon_g, \varepsilon_r, \varepsilon_i, \varepsilon_z)$ separately, using only its corresponding diffusion map and predictor variables. Starting with one of the bands, call it band $\band$, we optimise $\zeta_\text{Ia}$ over a grid of values of $\varepsilon_\band$. Then, for each value of $\varepsilon_\band$ in the grid, $\zeta_\text{Ia}$ is further optimized over a grid of proportions, $\gamma_\text{Ia}$. These optimisations are conducted using random forests of 500 trees and with the default value of $m_\text{try} = \lfloor \sqrt{m_\band} \rfloor$, where $m_\band$ is the number of predictor variables corresponding to the current band. To reduce over fitting, 
only the out of bag (OOB) predictions from the training set are used when computing (\ref{eq:zeta.Ia}). The OOB prediction of each SN (in the training set) is based only on the trees that were fit {\it without} that SN, i.e., trees for which that SN was not included in the bootstrap sample. In this way the OOB prediction is similar to a ``leave one out'' cross validation procedure.

In the second step, we fix $(\varepsilon_g, \varepsilon_r, \varepsilon_i, \varepsilon_z)$ at the values derived above and compute the final optimal values of $\gamma_\text{Ia}$ and $m_\text{try}$. In this step we use the combined predictor variables from all four bands.  Again we optimise $\zeta_\text{Ia}$ computed using OOB predictions from the training set, this time over a grid of values of $m_\text{try}$ and $\gamma_\text{Ia}$.

\section{Classification Results}\label{sec:results}

We now describe the classification results obtained using the two GP kernels described in Section \ref{sec:MeanAndCov}. We compare three GP kernel fits, given in Table \ref{tab:models}: the first  exclusively uses the SE kernel; the second exclusively uses the Gibbs kernel; and the third uses the Gibbs kernel whenever it improves the log posterior of the MAP estimate by more than 2\% over the SE kernel and uses the SE kernel otherwise. For each choice we train separately on the two training sets described in Section \ref{sec:data}, namely the biased training set, \Btrain, and the unbiased training set, \Utrain. In each case the results are evaluated on the appropriate test sets. We call the combined choice of GP kernel and training set a  ``model''.


\begin{table*}
	\centering
	\begin{tabular}{l c c}
		\hline \\[-1.8ex] 
		Designation & Kernel & Note \\ 
		\hline\hline \\[-1.8ex] 
		1 & SE & Applied to all LCs\\
		2 & Gibbs & Applied to all LCs\\
		3 & Gibbs / SE & Gibbs where log posterior increased by more than 2\%\\
		\hline
	\end{tabular}
	\caption{GP kernels compared in Sections~\ref{sec:results} and \ref{sec:staccato}.}
	\label{tab:models}
\end{table*}

\subsection{Tuning the Diffusion Map and Classifier}\label{sec:results.from.training}
\begin{table*}
	\centering
	\begin{tabular}{l c r r r r r r}
		\hline \\[-1.8ex] 
		Model & ($\varepsilon_g, \varepsilon_r, \varepsilon_i, \varepsilon_z$) & $m_\text{try}$& $\gamma_\text{Ia}$ & $\hat{I}_g$ & $\hat{I}_r$ & $\hat{I}_i$ & $\hat{I}_z$ \\ 
		\hline\hline \\[-1.8ex] 
		\bias{1} & $(10,2,10,20) \times 10^{-6}$ & 10 & 0.57 & 0.32 & 0.38 & 0.14 & 0.16 \\
		\bias{2} & $(10,2,10,10) \times 10^{-6}$ & 9 & 0.67 & 0.28 & 0.40 & 0.17 & 0.15 \\
		\bias{3} & $(10,2,20,20) \times 10^{-6}$ & 10 & 0.61 & 0.27 & 0.42 & 0.16 & 0.14 \\
		\\[-1.8ex] 
		\unbias{1} & $(2,2,2,6) \times 10^{-4}$ & 9 & 0.43 & 0.20 & 0.30 & 0.27 & 0.23 \\
		\unbias{2} & $(5,5,2,8) \times 10^{-4}$ & 9 & 0.44 & 0.20 & 0.26 & 0.29 & 0.25 \\
		\unbias{3} & $(3,2,2,2) \times 10^{-4}$ & 12 & 0.45 & 0.19 & 0.27 & 0.27 & 0.28 \\				
		\hline \\
	\end{tabular}
	\caption{Optimized tuning parameters for the diffusion maps and random forests (i.e., $\varepsilon_\band$ and $m_\text{try}$), classification threshold, $\gamma_\text{Ia}$, and relative importance of the different bands, ($\hat{I}_\band$, defined in Eq.~\eqref{def:importance_bands}), in the final combined random forests, for $\band\in \{g, r, i, z\}$. The model subscripts refer to the designations in Table~\ref{tab:models}.}
	\label{tab:tuning}
\end{table*}
Proceeding as described in Section~\ref{sec:tuning}, the optimal parameters for the diffusion maps and random forest classifier under each model appear in Table~\ref{tab:tuning}. In the table, we also show the relative weight of each band, quantified by 
\begin{equation} \label{def:importance_bands}
\hat{I}_\band = \frac{\sum_{v \in \mathcal{V}_\band} \hat{I}_v}{\sum_{v \in \mathcal{V}} \hat{I}_v}, \ \hbox{ for } \ \band \in \{g,r,i,z\},
\end{equation}
where $\mathcal{V}$ is the complete set of predictor variables generated by all four diffusion maps, $\mathcal{V}_\band$ is the subset of $\mathcal{V}$ corresponding to the variables generated by the diffusion map for colour band $\band$, and $\hat{I}_v$ is the ``importance'' associated with variable $v$. The importance of each variable is determined as the trees of the random forest grow.
Specifically, the decrease in gini index associated with each split is recorded alongside the variable that was divided into two regions in each partition step and the variable importance,  $\hat{I}_v$, is computed by separately summing the decreases in gini index associated with each variable \citep[p 593]{friedman2009}. 

\begin{figure*}
	\centering
	\includegraphics[width=\linewidth]{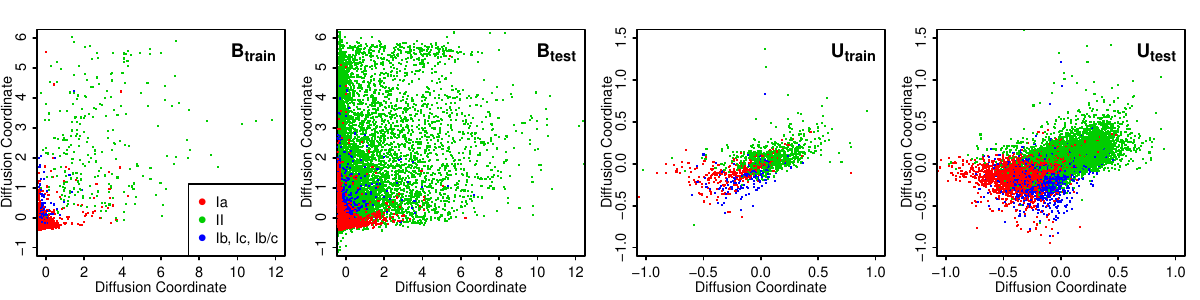}
	\caption{The two most important diffusion coordinates for models \bias{1} and \unbias{1}. \emph{Left Two Panels}: Diffusion coordinates for the test and training sets used with \bias{1}. \emph{Right Two Panels}: Diffusion coordinates for the test and training sets used with \unbias{1}. For the two test sets, the diffusion coordinates are computed after applying the Nystr{\"o}m extension of Eq.~(\ref{eq:nystrom}). The separation between SNIa and non-SNIa in \Utrain\ is much better maintained in \Utest\ than the separation in \Btrain\ is maintained in \Btest.}
	\label{fig31} 
\end{figure*}

The diffusion coordinates of the two most important variables as judged by the random forests are plotted and colour coded according to SN type for both the training and test sets under models \bias{1} and \unbias{1} in Figure~\ref{fig31}. (The other models are qualitatively similar.) While we plot only the two most important diffusion coordinates, we use up to 25 coordinates (in each band) in the classification, see Appendix~\ref{app:diff.maps} for details.
The separation between the three SN types is relatively good in \Btrain\ but less so in  \Btest. Although  \Utrain\ shows a higher degree of superposition between different types than does \Btrain, the separation that does exist for \Utrain\ does not degrade in \Utest.  We emphaise that the random forests are only trained to distinguish SNIa from non-SNIa and hence do not distinguish between sub-categories of non-SNIa (i.e., the green and blue dots in Figure~\ref{fig31}). 

\subsection{Classification Results}\label{sec:classification.results}
Classification results appear in Table~\ref{tab:results}. The first three columns summarise results for the training sets, where the optimality criterion, $\zeta_\text{Ia}$, the purity, $p_\text{Ia}$, and the efficiency, $e_\text{Ia}$, are estimated from the OOB samples. The last four columns summarize results for the corresponding test sets. The last column reports the area under the ROC (Receiver Operating Characteristic) curve, which is defined and discussed below. {Where appropriate, our results are compared with previous results in the literature in Table~\ref{tab:compare.results}.}

\begin{table*}
	\centering
	\begin{tabular}{llccccc}
		\hline
		& & \multicolumn{2}{c}{$\zeta_\text{Ia}$} & &\multicolumn{2}{c}{AUC}\\
        \cline{3-4}
        \cline{6-7}\\[-1.8ex]
		& Dataset		& Biased & Unbiased  && Biased & Unbiased \\
		\hline\hline \\[-1.8ex] 
		STACCATO 	& Original & 0.53	& 0.59	&& 0.96	& 0.98	\\
		\citet{kessler2010b}$^\dagger$ & Original & $0.3-0.45$ & -- && -- & -- \\ 
		\citet{Newling2011} & Original & 0.39 & -- && -- & -- \\ 
		\citet{Newling2011} & Updated & 0.15 & 0.45 && -- & -- \\ 
		\citet{Richards2012} & Updated & 0.13 & 0.31$^\star$ && -- & -- \\
        \citet{Varughese2015} & Original & 0.49 & 0.55&& -- & -- \\
		\citet{Lochner2016} & Updated & -- & -- && $\sim 0.88$ & 0.98 \\
		\hline
        		\multicolumn{7}{l}{$^\dagger$Summary of the performance achieved by the ``most stable'' (as a function of} \\	\multicolumn{7}{l}{redshift) classifiers entered in the original challenge.}\\
		\multicolumn{7}{l}{$^\star$\citet{Richards2012} used a resampling strategy that does not produce a truly}\\
        \multicolumn{7}{l}{\quad unbiased training set.}\\
	\end{tabular}
	\caption{Best results for various SNe classification methods. Although none of the entries are strictly comparable and comparisons should only be taken as a rough guide,  STACCATO results are superior to those obtained with previously published methods, as evidenced by its higher optimality criteria and AUC values.}
	\label{tab:compare.results}
\end{table*}

Models~\bias{1}--\bias{3} obtain high purity and efficiency values for \Btrain, with all of the purity values and two of the three efficiency values being above 95\%. The small differences among the models are likely due to the values of the threshold $\gamma_\text{Ia}$. In particular, Models \bias{2} and \bias{3} have higher values of $\gamma_\text{Ia}$ and higher purity, but lower efficiency.  Unfortunately, for all three models performance degrades substantially on \Btest, with purities ranging from 0.57 for model \bias{1} to 0.65 for model \bias{3} and efficiencies between of 0.80 and 0.88. Because poor purity is penalised harshly, $\zeta_\text{Ia}$ drops from $>0.86$ in \Btrain\ to $<0.31$ in \Btest. 

All of the models using \Utrain\ preform worse than their counterparts which use \Btrain\ in terms of both purity and efficiency on their respective training sets. This is not unexpected given the smaller separation of the SNIa and non-SNIa in \Utrain\ relative to \Btrain. (Compare the first and third panels of Figure~\ref{fig31}.) Unlike Models~\bias{1}--\bias{3}, however, performance does not degrade substantially when Models~\unbias{1}--\unbias{3} are applied to their test set.  If a classifier is tuned too precisely to the training set, its performance tends to degrade substantially in the test set. Using OOB samples aims to avoid this and appears to work quite well, at least with an unbiased training set. 

The degradation of the models~\bias{1}--\bias{3} when applied to \Btest\ illustrates a challenge that arises when the test set is not representative of the larger population. (This has been noted by others in this context, see \citet{Richards2012,Varughese2015,Lochner2016}) The good performance on the test set when the classifiers are trained on \Utrain, however, highlights the potential of our classification scheme. This gives us a guiding principle for developing an improved method: the bias in the training set must be addressed; this is the topic of Section~\ref{sec:staccato}.

\begin{table}
	\centering
    	\setlength\tabcolsep{4.5pt}
        \begin{tabular}{l r r r r r r r r r}
		\hline \\[-2.8ex] 
		& & \multicolumn{3}{c}{Training set} && \multicolumn{4}{c}{Test set} \\ 
        \cline{3-5} \cline{7-10} \\[-1.8ex]
		Model & $\gamma_\text{Ia}$ & $\tilde{\zeta}_\text{Ia}$ & $\tilde{p}_\text{Ia}$ & $\tilde{e}_\text{Ia}$ & &  $\zeta_\text{Ia}^*$ & $p_\text{Ia}^*$ & $e_\text{Ia}^*$ & AUC*\\
		\hline\hline \\[-1.8ex] 
		\bias{1}	& 0.57	& 0.88 & 0.97 & 0.97 && 0.27 & 0.57 & 0.88 & 0.93\\
		\bias{2} 	& 0.67	& $\mathbf{0.88}$ & 0.98 & 0.93 && 0.31 & 0.65 & 0.80 & 0.93\\
		\bias{3} 	& 0.61	& 0.87 & 0.97 & 0.95 && 0.28 & 0.59 & 0.85 & 0.93\\
		\\[-1.8ex]
		\unbias{1} 	& 0.43	& 0.60 & 0.88 & 0.86 && $\mathbf{0.59}$ & 0.87 & 0.86 & 0.98\\
		\unbias{2} 	& 0.44	& 0.56 & 0.88 & 0.80 && 0.59 & 0.88 & 0.83 & 0.98\\
		\unbias{3} 	& 0.45	& 0.58 & 0.88 & 0.84 && 0.58 & 0.87 & 0.84 & 0.98\\
		\hline \\[-1.8ex] 
		\multicolumn{10}{l}{$\tilde{}$ denotes OOB estimates from the training set.} \\
		\multicolumn{10}{l}{$^*$ based on predictions of the test set.}\\
	\end{tabular}
	\caption{Results of classifying SNIa using different models. Here $\zeta_\text{Ia}$ is the training criterion (\ref{eq:zeta.Ia}) with $W=3$,  $p_\text{Ia}$ is the purity, $e_\text{Ia}$ the efficiency, and AUC is the area under the ROC curve. The best results on $\zeta_\text{Ia}$ in the training sets and test sets are highlighted in bold. {The classification threshold, $\gamma_\text{Ia}$, from Table \ref{tab:tuning} is also shown, to illustrate its influence on the three classification measures.}}
	\label{tab:results}
\end{table}

\begin{figure}
	\centering
	\includegraphics[width=\linewidth]{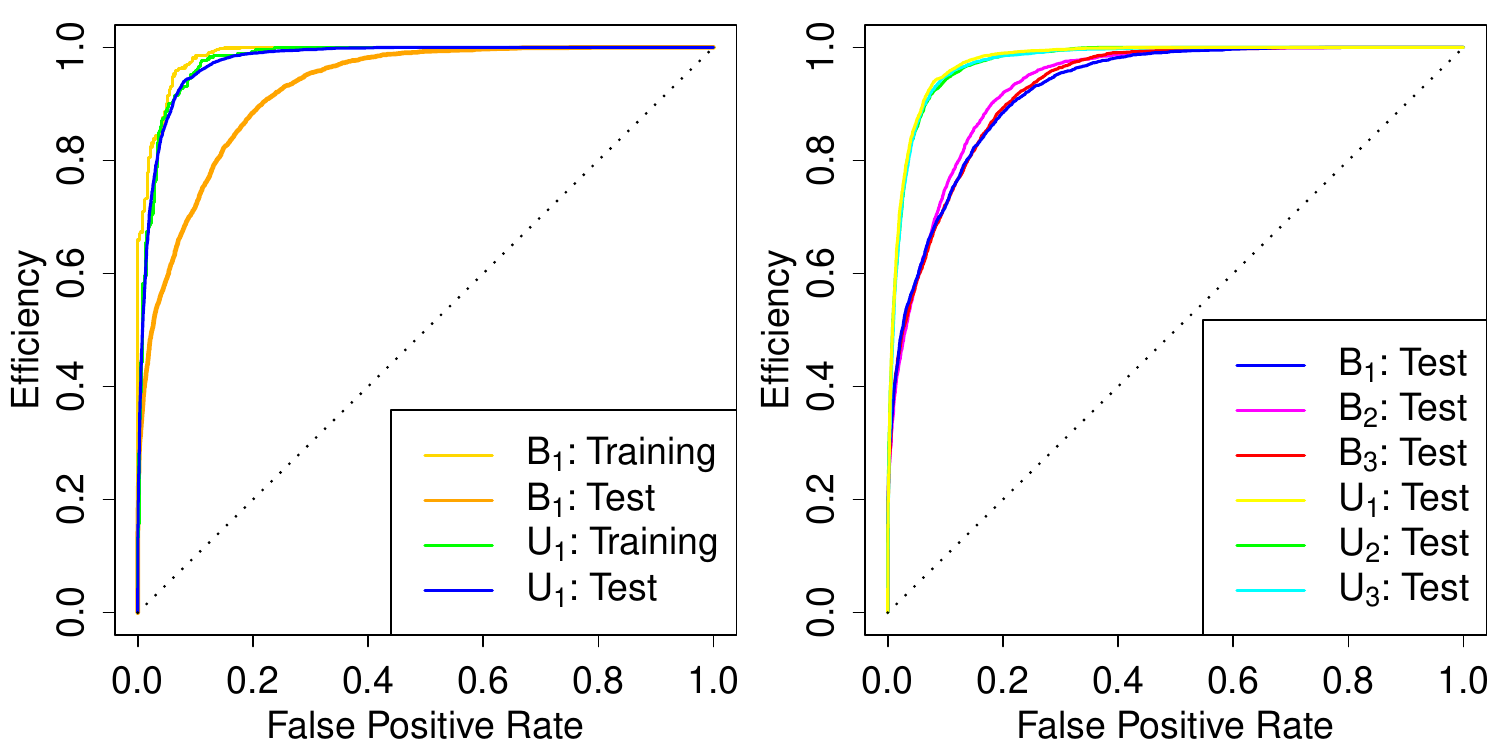}
	\caption{\emph{Left}: ROC curves for models \bias{1} and \unbias{1} on both training and test sets. \emph{Right}: ROC curves for all models on test data only.}
	\label{fig34}
\end{figure}

Table \ref{tab:results} compares the performance of each model using a single classification threshold, $\gamma_\text{Ia}$, for each. To better explore the trade-off between purity and efficiency, we can plot efficiency against the false positive rate (the number of false positives relative to the total number of negatives, i.e. non-Ia) of the predictions using a fine grid of $\gamma_\text{Ia}$ values between 0 and 1. Such plots are known as Receiver Operator Characteristic (ROC) curves and appear in Figure~\ref{fig34}. A perfect classifier would have an efficiency of one and a false positive rate of zero and would appear in the upper left corner of the panels in Figure~\ref{fig34}. The ROC curve of a random classifier, on the other hand, would appear as a 45 degree line, as shown as a dotted black line in Figure~\ref{fig34}. The area under the ROC curve (AUC) is a simple summary of overall performance. A perfect classifier would have an AUC of 1, while a random classifier would have an  AUC of 0.5. 

As expected, the ROC curves for \Utest\ are above those for \Btest, indicating better performance. Likewise the AUC values in Table~\ref{tab:results} indicate better performance with \Utest. Comparing both the ROC curves and the AUC values for the three different GP kernels, however, shows no substantial difference between the GP models. Thus, the differences in the kernel LC fits (e.g., in Figure~\ref{fig16}) somewhat surprisingly do not translate to differences in classification performance. 

\begin{figure}
	\centering
	\includegraphics[width=\linewidth]{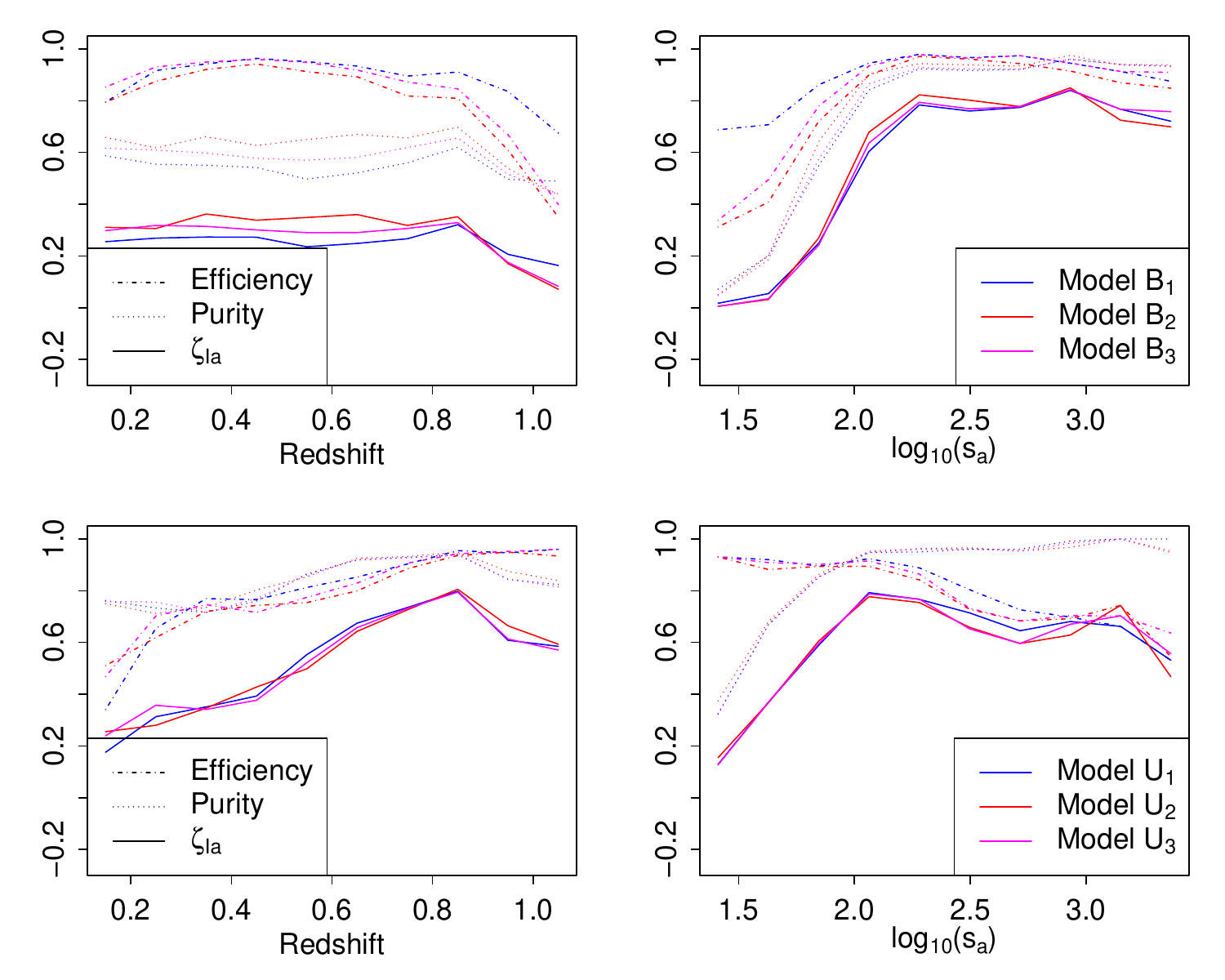}
	\caption{\emph{Top row}: Performance of models trained on \Btrain\ when used to classify \Btest, as a function of redshift and as a of function of $s_a$, the log brightnesses. \emph{Bottom row}: Same but for models trained on \Utrain\ and tested on \Utest.}
	\label{fig39}
\end{figure}

Figure~\ref{fig39} plots the performance of the models as a function of redshift (left panels) and as a function of the log brightnesses, $\log(s_a)$ (right panels). (Recall that $\log(s_a)$ is a measure of apparent brightness for SN $a$, see (\ref{eq:normalising.flux}).) 
The classifiers trained on \Btrain\ perform equally well on \Btest\ at all redshifts up to $z \lesssim 0.9$, but  efficiency and purity drop for larger $z$ (top left panel). For classifiers trained in \Utrain,  performance on \Utest\ increases to $z \approx 0.9$, mainly due to increasing efficiency, and then  performance drops at the highest redshift values (bottom left panel). The efficiency is generally lower than purity with $\mathcal{U}$, and vice-versa with $\mathcal{B}$. This can be attributed to the different classification thresholds, $\gamma_\text{Ia}$, which in turn is due to the biased composition of \Btrain.

As a function of brightness the classifiers trained on \Btrain\ perform well on the brighter SNe (larger values of $\log s_a$).  For SNe with $\log s_a > 5$, both the purity and efficiency for the best model are above 90\%. However, with fainter SNe the performance is much poorer with very low purity levels. For the models trained on \Utrain\ the same qualitative description holds, although efficiency is higher for fainter SNe. 

Finally, we check whether the number of observation times in the LCs influences the performance of the classifiers. Surprisingly classification is equally good with SNe with many observations and with those with only few. 

The overall picture emerging from this analysis is that the classifiers struggle with dimmer, higher-redshift SNe, more so when trained with \Btrain. To improve performance, we must account for the bias in the training set and improve training for high redshift/low brightness SNe. This is the guiding idea behind our STACCATO method, which we introduce in the next section.

\section{Improving the Classifier with STACCATO}\label{sec:staccato}

The SN classification challenge of \citet{kessler2010} provided a biased training set because this is what is expected in a real sample of spectroscopically confirmed SNe. {Dealing with the issue of selection bias is a common problem in astronomy, and it was recognized in the early 20th century. Today, astronomers often refer to \cite{1925MeLuF.106....1M} bias to indicate truncation in magnitude-limited samples; and to~\citet{1913MNRAS..73..359E} bias to indicate the effect of measurement error near the sensitivity limit of a survey, when the underlying sources have a steep distribution of the latent property one is trying to measure. The usual way to address selection bias is by ``correcting'' for it via simulation (in cosmological SN analysis, see e.g.~\cite{Kessler:2016uwi}), although arguably better Bayesian solutions exist (e.g., \cite{Budavari:2008sw,Rubin:2015rza,Kelly:2007jy,gelman2013}). Here, we are interested in the specific issue of how to overcome the effect of selection bias in the training sample, which thus become non-representative of the full (test) sample. In other contexts, data augmentation schemes have been proposed to address this problem (e.g., \cite{Hoyle:2015ada}). In this section, we present a new method that aims to improve the performance of classifiers that use only a biased training set, \Btrain.}
In particular, in Section \ref{sec:propensity.score} we use propensity scores to partition \Btrain\ and then suggest a novel approach in Section~\ref{sec:synthetic} that uses the partition to improve classification. We call our approach ``Synthetically Augmented Light Curve Classification'' or STACCATO. Given the similarities between classifiers based on the GP kernels considered in Section~\ref{sec:classification.results}, STACCATO is implemented only with the SE kernel and only with \Btrain. It could, however, be applied to any set of GP fitted LCs using any similarly biased training set.

\begin{figure}
	\centering
	\includegraphics[width=\linewidth]{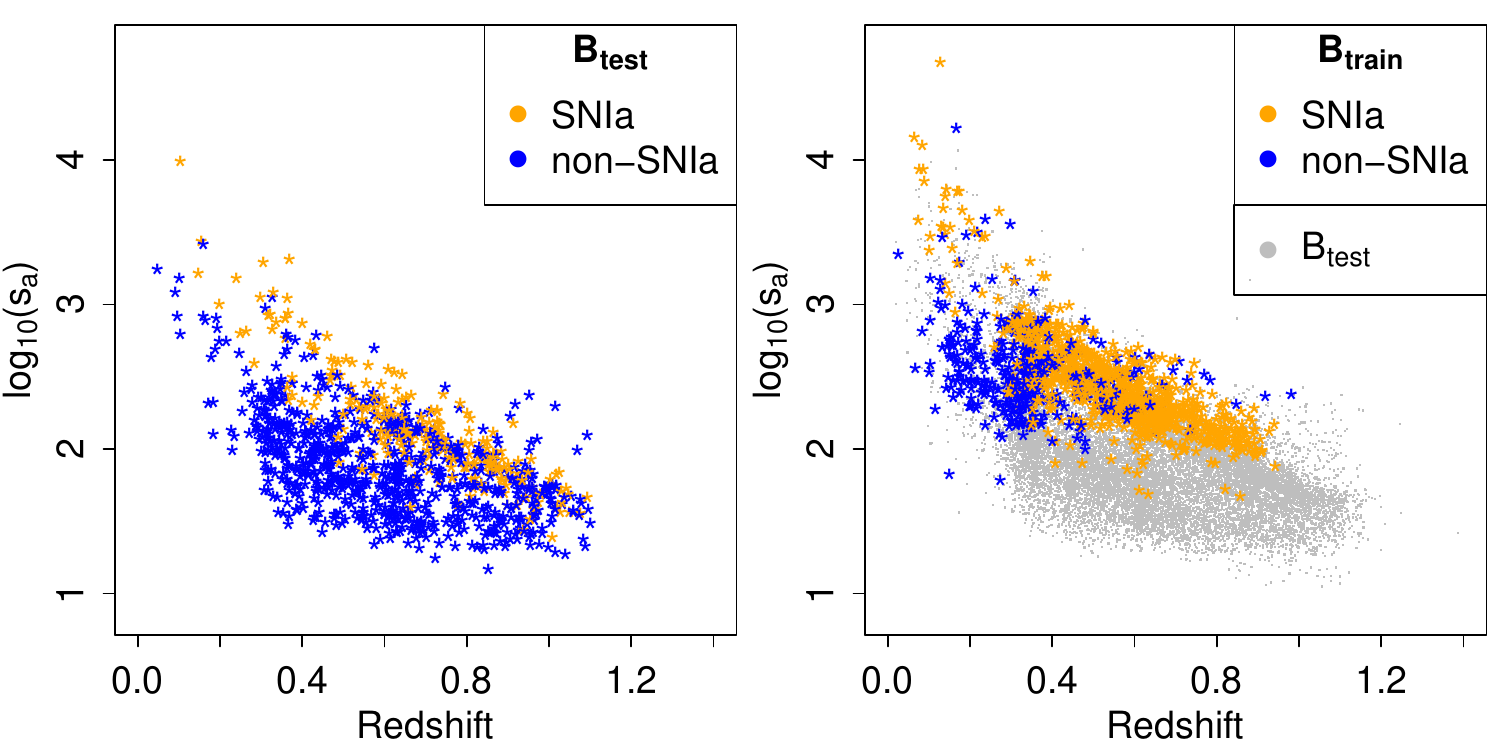}
	\caption{\emph{Left}: Scatterplot of $\log s_a$ against redshift for a random subset of 1,000 SNe from \Btest. \emph{Right}: Same plot for SNe in \Btrain. Both panels are colour coded by SN type. In the right panel the grey dots represent the test data for better comparison of the domains.}
	\label{fig38}
\end{figure}

\begin{figure}
	\centering
	\includegraphics[width=\linewidth]{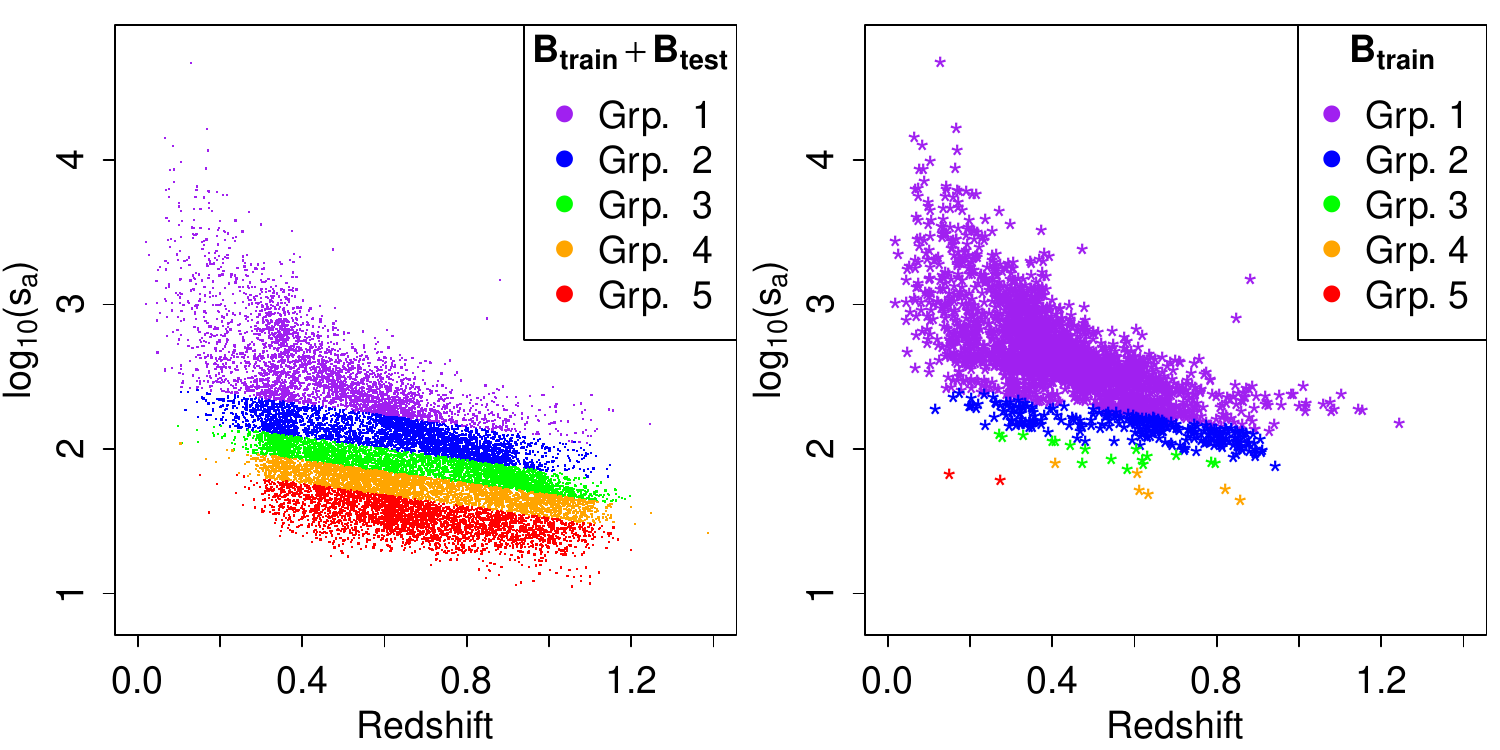}
	\caption{Propensity Score Groups. \emph{Left}: Scatterplot of $\log_{10} s_a$ against redshift for the entire dataset (\Btrain + \Btest), colour-coded according to the five groups in Table \ref{tab:prop.group}, obtained by partitioning on the fitted propensity score. \emph{Right}: The same, but showing only the data from \Btrain. One can clearly see the paucity of training data in groups 3--5.}
	\label{fig44}
\end{figure}

\subsection{Grouping SNe by Propensity Scores}\label{sec:propensity.score}

Figure~\ref{fig38} illustrates the excess of bright SNe in \Btrain, especially at high redshift. The left panel shows a scatter plot of redshift and brightness (in terms of $\log s_a$, from Eq.~\ref{eq:normalising.flux}) for a random sample of \Btest; the right panel plots the same variables for \Btrain.  Clearly, \Btrain\ over-represents brighter SNe and completely lacks the fainter SNe found in \Btest, especially at high redshift. 

Propensity scores are used in observational studies to quantify the probability that a unit is in the treatment group rather than the control group \citep{rosenbaum1984}. Here we define the propensity score as the probability that a particular SN belongs to \Btrain, as function of a set of observed covariates. We can partition the data based on their propensity scores into a number of groups, such that within these groups the bias between the training and test set is substantially reduced. \citet{rosenbaum1984} argue that (under certain assumptions) five groups are sufficient to remove approximately 90\% of the bias. Propensity scores can naturally be modelled using a logistic regression with the response variable being membership in \Btrain . Ideally all covariates that exhibit bias (i.e., a different distribution in \Btrain\ and \Btest) should be included in the regression so as to balance their distributions in the groups. For illustration we include redshift and brightness in the logistic regression and define the propensity score to be the conditional probability of a randomly selected SN being a member of \Btrain\ given its brightness and redshift. Table~\ref{tab:log.reg} reports the fitted (maximum likelihood) coefficients for the logistic regression. 

\begin{table}
	\centering
	\begin{tabular}{lrrr}
		\hline
		Predictor variable & \multicolumn{1}{c}{Estimate} &  p-value \\ 
		\hline\hline \\[-1.8ex] 
		(Intercept) & $-10.6790 \pm 0.3348$ & $<10^{-4}$  \\ 
		Redshift & $1.3412 \pm  0.2009$ &  $<10^{-4}$  \\ 
		Brightness & $1.4751 \pm  0.0466$ &  $<10^{-4}$  \\ 
		\hline
	\end{tabular}
	\caption{Logistic regression applied to the entire dataset ($n$=17,330). The response  variable identifies which SNe are in \Btrain\ $(=1)$ and which are in \Btest\  $(=0)$. (Apparent) Brightness is described by $\log(s_a)$ from Eq.~(\ref{eq:normalising.flux}). We also show the p-value for including each predictor variable in the model, given the others.}
	\label{tab:log.reg}
\end{table}

Following \citet{rosenbaum1984}, we use the quintiles of the estimated propensity scores (i.e., the fitted values from the logistic regression) to split the entire data set into five equally-sized groups. The resulting partition into groups is plotted in Figure~\ref{fig44} and summarised numerically in Table~\ref{tab:prop.group}. Due to the severe bias in brightness, Groups~3--5 contain very few SNe from the training set, \Btrain: specifically, they contain just 17, 6, and 2 of the 1,217 SNe in \Btrain. Notice that even though the SNe types are not used in the linear regression, the type imbalances between the test and training sets are greatly reduced in Groups~1--2 compared to the imbalance in the test and training sets as a whole (cf. Table \ref{tab:datasets}).

\begin{table}
	\centering
	\begin{tabular}{clrrr}
		\hline
        && Number & Number & Proportion\\
		Group & Set & \multicolumn{1}{c}{of SNe} 
        	& \multicolumn{1}{c}{of \SN} & \multicolumn{1}{c}{of \SN} \\ 
		\hline\hline \\[-1.8ex] 
		1 & Training & 947  & 652  & 0.69\\
		& Test     & 2519 & 1242 & 0.49\\ \hline
		\\[-2.4ex]
		2 & Training & 245  & 181  & 0.74\\
		& Test     & 3221 & 1147 & 0.36\\ \hline
		\\[-2.4ex]
		3 & Training & 17   & 12   & 0.71\\
		& Test     & 3449 & 754  & 0.22\\ \hline
		\\[-2.4ex]
		4 & Training & 6    & 6    & 1\\
		& Test     & 3460 & 342  & 0.10\\ \hline
		\\[-2.4ex]
		5 & Training & 2    & 0    & 0\\
		& Test     & 3464 & 107  & 0.03\\   
		\hline
	\end{tabular}
	\caption{Composition of the five groups obtained by partitioning the data based on the fitted propensity scores.}
	\label{tab:prop.group}
\end{table}

\begin{figure*}
	\centering
	\includegraphics[width=\linewidth]{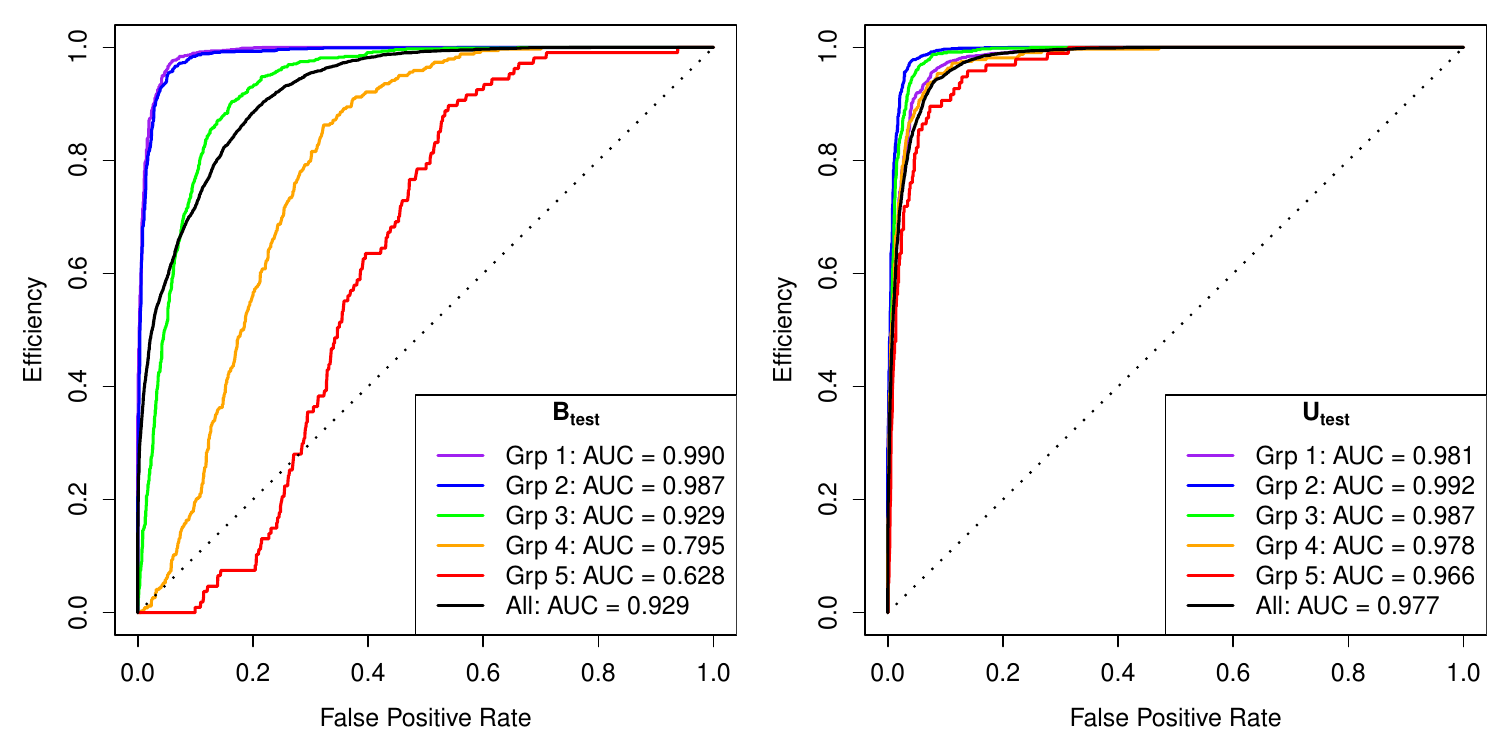}
	\caption{ROC curves for the {five test groups} in Table~\ref{tab:prop.group}, computed using a classifier trained with the biased training set (right) and with the unbiased training set (left). In both cases the grouping is only used for partitioning the test set. The random forest classifier is fit and tuned using the combined data.}
	\label{fig41}
\end{figure*}

In Figure~\ref{fig41} ROC curves for models \bias{1} (left) and \unbias{1} (right) are plotted for the five propensity score groups from Table~\ref{tab:prop.group}, where the grouping is only used for partitioning the test set; the random forest classifier is fit and tuned using the combined data. Under model~\bias{1}, the ROC curves become coarser with increasing group number because of the smaller numbers of SNIa in the Groups~3--5. The performance in Groups~1--2, where a large proportion of \Btrain\ resides, is very good with almost perfect classification. Performance degrades in Groups~3--5, however, to the degree that classification in Group~5 is not much better than a random classifier. 

When using an unbiased training set (right panel of Figure~\ref{fig41}) classification is nearly equally good in all five groups, with only slightly worse classification in Group~5. This recalls previous results that conclude that an unbiased training set is necessary to obtain good performance with standard classification methods~\citep{Lochner2016,Varughese2015}. In the next section we show how propensity scores can be used to obtain good classification even with a biased training set.   

\subsection{STACCATO: Synthetically Augmented Light Curve Classification}\label{sec:synthetic}

\subsubsection{STACCATO methods}\label{sec:staccato-methods}
Our approach to mitigate bias and improve LC classification is to train the classifier separately within the five groups defined by the propensity scores. {Thus we divide both the test and training set into five groups according to the propensity scores to obtain the five {\it test groups} and five {\it training groups}, respectively} (See Table~\ref{tbl:terms} for group terminology.) An immediate difficulty  {with obtaining group-specific classifiers}
is the lack of data in {Training Groups~3--5}, and in particular the lack of SNIa. In situations with imbalanced data in one or both categories, a popular strategy is to over-sample the underrepresented observations and/or to down sample the over-represented observations \citep{japkowicz2000}. Rather than resampling {the sparse training groups (corresponding to Groups~3--5)}, we create {\it augmented training {groups}} by synthesizing 
observations in the LC space, a procedure inspired by \citet{ha1997,chawla2002} -- an approach we call `STACCATO'.

\begin{table}
{
	\centering
	\begin{tabular}{lp{4.5cm}}
		\hline
        Term & Meaning \\
        \hline\hline
        Groups & The entire dataset is split into five equal sized groups based on the propensity scores.\\[4pt]
        \parbox[t]{3cm}{Test (Training)\\ Groups} &  SNe in the test (training) set are partitioned according to their group membership.\\[4pt]
        \parbox[t]{3cm}{Augmented\\ Training Groups} &  The training groups are augmented with other training groups and/or synthetic LCs before being used to train the group-specific classifiers.\\[4pt]
        \parbox[t]{3cm}{Validation and\\ Generalization Groups} & Each test group is randomly divided into a validation and generalization group, such that each validation groups contains 1500 SNe.\\
		\hline
	\end{tabular}
	\caption{Terminology used for group-specific analyses.}
	\label{tbl:terms}
    }
\end{table}

We thus exploit the probabilistic nature of the GPs to over-sample the LCs by sampling `synthetic' curves from the fitted GPs. More precisely, we augment the {small training groups} by generating  $K$ synthetic LCs for each SN in that group, where the {\it augmentation factor}, $K$, may vary between {training} groups and where each synthetic LC is drawn band-by-band from the fitted GP for that SN.
Compared with simply resampling {Training} Groups 3--5, our augmented training {groups} have two advantages. First, the posterior mean vectors of the GPs that we use as the input for the diffusion maps are only estimates and may differ appreciably from the true LCs, at least with noisy data. By sampling LCs from the GPs, the augmented training {group} takes these uncertainties into account. Second, sampling additional LCs creates a richer sample in the regions of diffusion space most relevant for classification.

\begin{table*}
	\centering
	\begin{tabular}{cclclclclcl}
		\hline
		Test & \multicolumn{10}{c}{{Training Groups}}\\
		{Group} & \multicolumn{2}{c}{1} & \multicolumn{2}{c}{2} & \multicolumn{2}{c}{3} & \multicolumn{2}{c}{4} & \multicolumn{2}{c}{5} \\ 
		\hline\hline \\[-1.8ex] 
		1 & + & (0) & +/--& (0-2) & -- & & -- && --\\
		2 & +/-- & (0) & + &(0-2) & -- & & -- && --\\
		3 & -- & &  +/-- & (0)   & + &(0-10)  & + &(0-10) & + &(0-10)\\
		4 & -- & &  +/-- & (0)   & + &(0-10)  & + &(0-10) & + &(0-10)\\
		5 & -- & &  +/-- & (0)   & + &(0-10)  & + &(0-10) & + &(0-10)\\
		\hline
	\end{tabular}
	\caption{{Composition of the augmented training {groups} (along rows) for each test {group} (rows).} A `+' indicates that an augmented training {group} included the original training {group} corresponding to a particular column; a `--' indicates that original training {group} is not used; a `+/--' indicates that both combinations were tried. The numbers in parentheses give the number of LCs synthesized for each of the LCs in the original training {group}, i.e. the augmentation level. {Thus a `+ (0)' symbol means that the original training {group} was used, with no augmentation.}}
	\label{tab:training.composition}
\end{table*}

The key innovation of STACCATO is to construct an augmented training {group} for each of the {five test groups}. The augmented training {group} for each {test group} may be constructed using {more than one of the original training groups}, and each may be supplemented with synthetic LCs using different augmentation factors.  As there are many ways to augment the training {groups}, choosing the augmentation scheme may benefit from an optimization step. The rows of Table \ref{tab:training.composition} give the augmented training {groups} configurations considered for each of the five {test groups}. Here, a `+' indicates that the {SNe of a given training group (columns) are included in the augmented training set for that test group (rows)}; a  `--' indicates that they are not included;  a `+/--' indicates that both possibilities are considered; and the numbers in parentheses are the considered augmentation factors, i.e., 
the number of synthetic LCs that are sampled for each {SN of the original training group}. While this is not an exhaustive search of the possible configurations, it is meant to span a wide range of choices. We then select the optimal configuration as the one that gives the highest AUC within each {test group}, as described below.

\begin{table*}
	\centering
	\begin{tabular}{clllllc c | c}
		\hline
		Test & \multicolumn{5}{c}{Optimal {training group} configuration} & AUC with & AUC w/o & AUC\\
		{Group} & 1 & 2 & 3 & 4 & 5 & synthetic LCs  &  synthetic LCs & original\\ 
		\hline\hline 
		\\[-1.8ex] 
		1 & + (0)& --& -- & -- & -- 		& 0.991 & 0.991 & 0.993\\
		2 & + (0)& + (2)& -- & -- & -- 		& 0.989 & 0.990 & 0.988\\		
		3 & --& + (0)& + (0)& + (5)& + (5)	& 0.968 & 0.958 & 0.926\\
		4 & --& + (0)& + (5)& + (10)& + (5)	& 0.919 & 0.887 & 0.791\\
		5 & --& + (0)& + (6)& + (10)& + (2)	& 0.842 & 0.709 & 0.636\\
		\hline
	\end{tabular}
	\caption{Optimal configuration for STACCATO classification along with their AUCs. For comparisons the AUC of the same configuration but without adding synthetic LCs {(i.e., corresponding to setting all numbers in the parentheses to zero)} and the AUC using the original \Btrain, computed without STACCATO are also given. (The original analysis is the same as summarized in Figure~\ref{fig41}. Each AUC is computed using only the corresponding generalization {group}, resulting in small variations in the last column compared with Figure~\ref{fig41}). Notation is the same as in Table~\ref{tab:training.composition}.}
	\label{tab:optimal.composition}
\end{table*}

The sampled synthetic LCs are standardized in exactly the same manner as the original data. For each possible configuration of the augmented training {groups}, the diffusion maps are constructed using only the LCs in that configuration. For simplicity we set $\epsilon=2\times 10^{-5}$ in all diffusion maps and $m_\text{try} = \lfloor \sqrt{\sum_{\band  \in C} m_\band} \rfloor$ in all random forests, where $C=\{g,r,i,z\}$. (I.e., the diffusion maps and random forests are {\em not} optimized over $\epsilon$ and $m_\text{try}$ as described in Section \ref{sec:tuning}.)  For each of the five groups, we want to select the configuration that gives the highest AUC within that group. In order to accomplish this, we follow \citet[ch. 7]{friedman2009} and partition the test {group} into a validation and a generalisation {group}, 
with $1500$ LCs in {each validation group}. The validation {groups} are used to compute the AUC for each configuration, and the configuration that leads to the highest AUC is chosen. The generalization {groups} are used to measure the performance of the best configuration. 
In practice, the SN types in the test {groups} are not known so the training {groups} or subsets of the training {groups} should instead be used to compare the AUC for each configuration. This, however, may be infeasible in practice when the training {groups} are very small (e.g., Training Groups~3--5). Alternatively, the optimal configuration choice can be determined using a simulation study, such as that used in the SN photometric classification challenge.

\subsubsection{STACCATO summary}\label{sec:staccato-summary}

In summary, STACCATO proceeds by

\begin{enumerate}
\item Fitting the LCs using a GP implemented with a squared exponential kernel as described in Section~\ref{sec:modelling};
\item {Forming the training groups} by grouping the training set according to a fitted propensity score model as described in Section~\ref{sec:propensity.score};
\item Augmenting each of the {training groups} with observed LCs from  {other training groups} and/or synthetic LCs sampled under the GP fits of observed SNe, as described in Section~\ref{sec:staccato-methods};
\item Time-aligning and normalizing the LCs as described in Section~\ref{sec:norm}; 
\item Separately computing diffusion maps for each of the augmented training {groups}, including the Nystr\"om extension described in Appendix~\ref{subsec:Nystrom};
\item Classifying the SNe in each {training} group using a separate random forest as described in Section~\ref{sec:random.forest}.
\end{enumerate}

Step (iii) requires an augmentation scheme. In our current implementation each scheme is determined via recursive optimization over a validation {group}. Other schemes that do not require validation {groups} are possible and are the subject of ongoing research. 

\subsubsection{STACCATO results}\label{sec:staccato-resuts}

Table~\ref{tab:optimal.composition} lists the optimal configurations (determined by optimizing the AUC on the validation {groups}), along with their AUCs, computed on the generalisation {groups}. For comparison we also give the AUCs of the same configuration but without adding the synthetic LCs to the augmented training {groups}. (The training {groups} are still augmented in that they may combine {more than one of the original training groups}.) 
We also report the AUCs using {each of} the generalization {groups} without any augmentation. (Although this analysis follows the same steps as that illustrated in Figure~\ref{fig41}, the AUC values are slightly different than those reported in  Figure~\ref{fig41} because the original analysis used the entire test set rather than just the generalization {groups}.)
STACCATO delivers significantly larger AUC across all {test groups}, particularly when synthetic LCs are included in the augmented training {groups}.

\begin{figure}
	\centering
	\includegraphics[width=\linewidth]{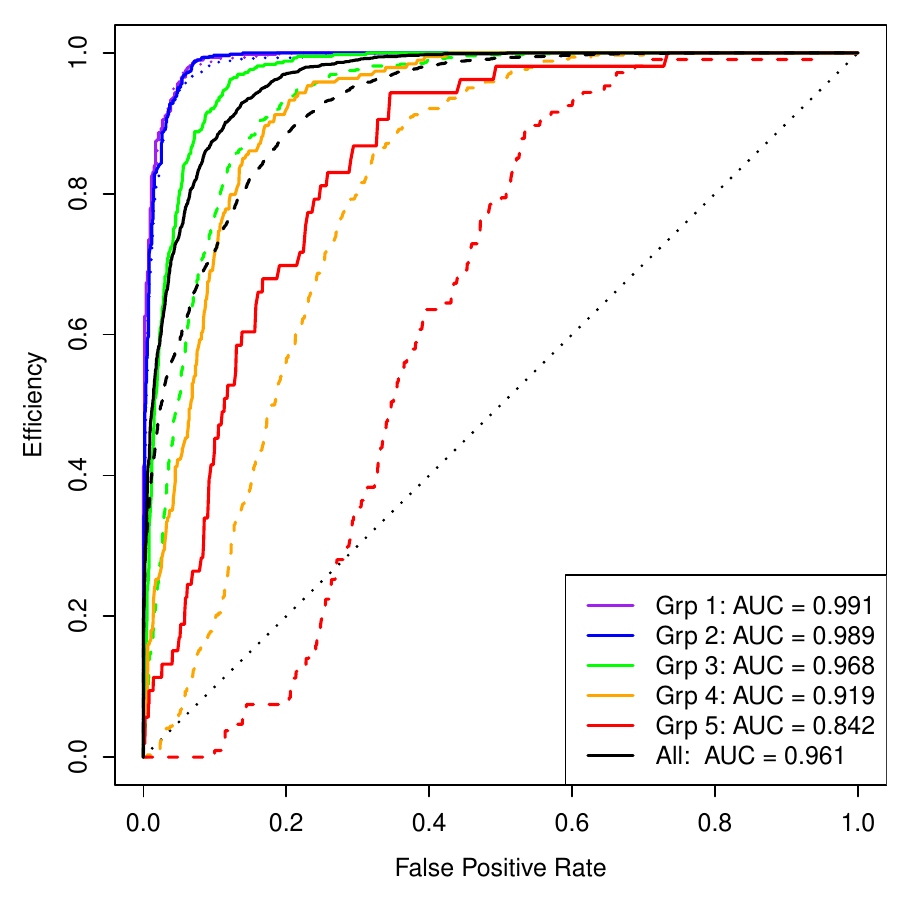}
	\caption{Classification with STACCATO. ROC curves of the optimal configurations given in Table~\ref{tab:optimal.composition}. The curves are computed using the generalization {groups}. For comparison, ROC curves from Figure \ref{fig41} (without STACCATO) are plotted as dashed curves.}
	\label{fig42}
\end{figure}

Figure~\ref{fig42} compares the ROC curves for the optimal configurations given in Table~\ref{tab:optimal.composition} (i.e., using the augmented training {groups} with synthetic LCs) with those given in Figure~\ref{fig41}. The substantial classification improvements apparent for Groups~3--5 are largely driven by the addition of synthetic LCs. The optimal training {group} for Group~5, the group that benefits the most from augmentation, includes 436 LCs, of which $\approx 38\%$ are synthetic. In contrast, Groups~1 and 2 show no improvement when synthetic LCs are introduced.
Combining classification among all five groups, the AUC increases from 0.919 in the original model to 0.961 with STACCATO. Compared with the AUC of 0.977 obtained when using the unbiased training set, STACCATO performs almost as well as this gold standard. 

{
Ultimately, we propose to use STACCATO to compute the probability that any particular photometrically observed SN is a SNIa. These probabilities can be used as weights or priors in secondary analyses and and their computation does not require artificial hard thresholding. For comparison with existing methods, however, we propose a dynamic threshold that increases with the propensity group number and thus accounts for 
the fact that there are more SNIa among the bright SNe (low numbered groups).
Specifically,} setting arbitrary classification thresholds of $\gamma_\text{Ia} = 0.50, 0.60, 0.70, 0.80, 0.90$ for Groups~1--5, we obtain an efficiency of $e_\text{Ia}=0.846$, a purity of $p_\text{Ia}=0.834$ and an optimality criterion of $\zeta_\text{Ia}=0.529$ on the generalisation {groups}. Even without any optimization of the classification thresholds, we achieve a better optimality criterion than any of the previous results on the same data set (see Table \ref{tab:compare.results}). We therefore simply adopt our arbitrary choice for the classification thresholds, without any form of optimization on the threshold, to further compare STACCATO results, obtained both with  \Utrain\ and \Btrain, with other published methods in Table~\ref{tab:compare.results}. 

The comparison is complicated by the fact that different authors used different version of the SN classification challenge simulations. \citet{Richards2012} and \cite{Lochner2016} adopted an updated version of the simulated data with a number of corrections and bug fixes that was released after the challenge \citep{kessler2010b}. \citet{Newling2011} report that classification using the updated simulation is more difficult. This might be in part due to the fact that two bugs where discovered in the original simulation (the one used in this paper), which made classification easier: some SNIa were too bright, while {\em all} non-SNIa were too dim by a factor $(1+z)$~\citep{kessler2010b} (We will apply STACCATO to the newer, more difficult simulation in a future paper.)
Further, the results of the original challenge are presented as a function of redshift, with no combined result reported, and only \Btrain\ was considered.  \citet[p. 8]{kessler2010b} note that the ``most stable'' classifiers obtained  $\zeta_\text{Ia} \in [0.3,0.45]$ at all redshifts. (Some classifiers achieved $\zeta_\text{Ia} = 0.6$ at some redshift, but with large downwards variations of a factor of 2 at other redshift values, suggesting an overall poorer performance once averaged over redshift.) Although none of the entries in Table~\ref{tab:results} are exactly comparable, 
results obtained with STACCATO appear to be superior to all previously published methods, as evidenced by the its higher optimality criterion and AUC values.

\section{Discussion and Conclusions}
\label{sec:conclusions}

Augmenting a biased training set with synthetic LCs simulated from fitted GPs can dramatically improve SN classification, without needing to observationally obtain an expensive unbiased training set. To our knowledge, this is the first demonstration of this technique. Although the quality of the GP LC models vary when judged by eye, we find that classification results are nonetheless insensitive to the choice of GP covariance function.

We achieve increased performance levels in AUC, from 0.93 using the biased training set to 0.96 using STACCATO. This compares well with the best result in the literature, an AUC of 0.88,  obtained by~\cite{Lochner2016} with a biased training set. STACCATO also compares favourably with the gold standard of 0.98 obtained using our method with an unbiased training set, which itself matches the performance of previous methods applied with representative training sets \citep{Lochner2016}. We also obtain significantly higher purity values than previous studies with the same simulated data (0.59, compared to the previous best result of 0.55 obtained by ~\citet{Varughese2015}).

Although \citet{Moller:2016vih} obtain an AUC of 0.98 by applying random forests and boosted decision trees to a simulated SNLS3 SN sample, this simulation is explicitly designed to include a larger number of fainter objects than the standard training set. This effectively creates a representative training set.  Indeed they find that a set of specially designed selection cuts is ``essential'' for good classification. Our approach demonstrates that such cuts can be avoided. 

\citet{Dai:2017loy} also report an AUC of 0.98 on a set of LSST-like SN simulations using a random forest classifier.  However, this result requires a representative training set that is 30\% the size of the test set. As \citet{Dai:2017loy} acknowledge, this is impossible to obtain in practice, hence they suggest using a simulation to generate a large representative training set. Again, STACCATO achieves comparable performance without needing such a large representative training set. 

Another advantage of STACCATO is that the propensity scores groups can be used to assess the classification quality. In propensity scores groups with a small training set, accurate classification may not be possible, while it can be very accurate in groups with a large training set. Furthermore, STACCATO enables the identification of a subset of SNe for spectroscopic follow up with a high probability of containing a sizable proportion of SNIa. For example, among the 20 SNe in Group~5 that are most likely to be SNIa (as judged by random forest votes) five are actually SNIa, a proportion of 25\%. Given that only 2.7\% of the SNe in Group~5 are SNIa, this is a considerable improvement in the proportion of SNIa among those identified for follow-up. For comparison, when using the biased training set, only one SNIa is found among the 200 most likely SNe in the same group, a fraction of 0.5\%. 
Thus STACCATO leads to a 50 fold improvement in the probability
that a SN identified for spectroscopic follow-up at high redshift is a SNIa (i.e., the true positive rate).

The propensity score groups have differing proportions of SNIa even after controlling for redshift and brightness, which suggests that there are other variables that could be informative in modeling the inclusion of SNe in the training set (e.g., via logistic regression).  It would be interesting to include additional variables (such as the signal to noice ratio, data quality cuts, etc) that bias spectroscopic follow-up in favour of SNIa. This would enable the propensity scores to identify groups where the training set better represents the test set in terms of the distributions of the additional variables in the training and test sets.   

\begin{figure}
 	\centering
 	\includegraphics[width=\customwidth]{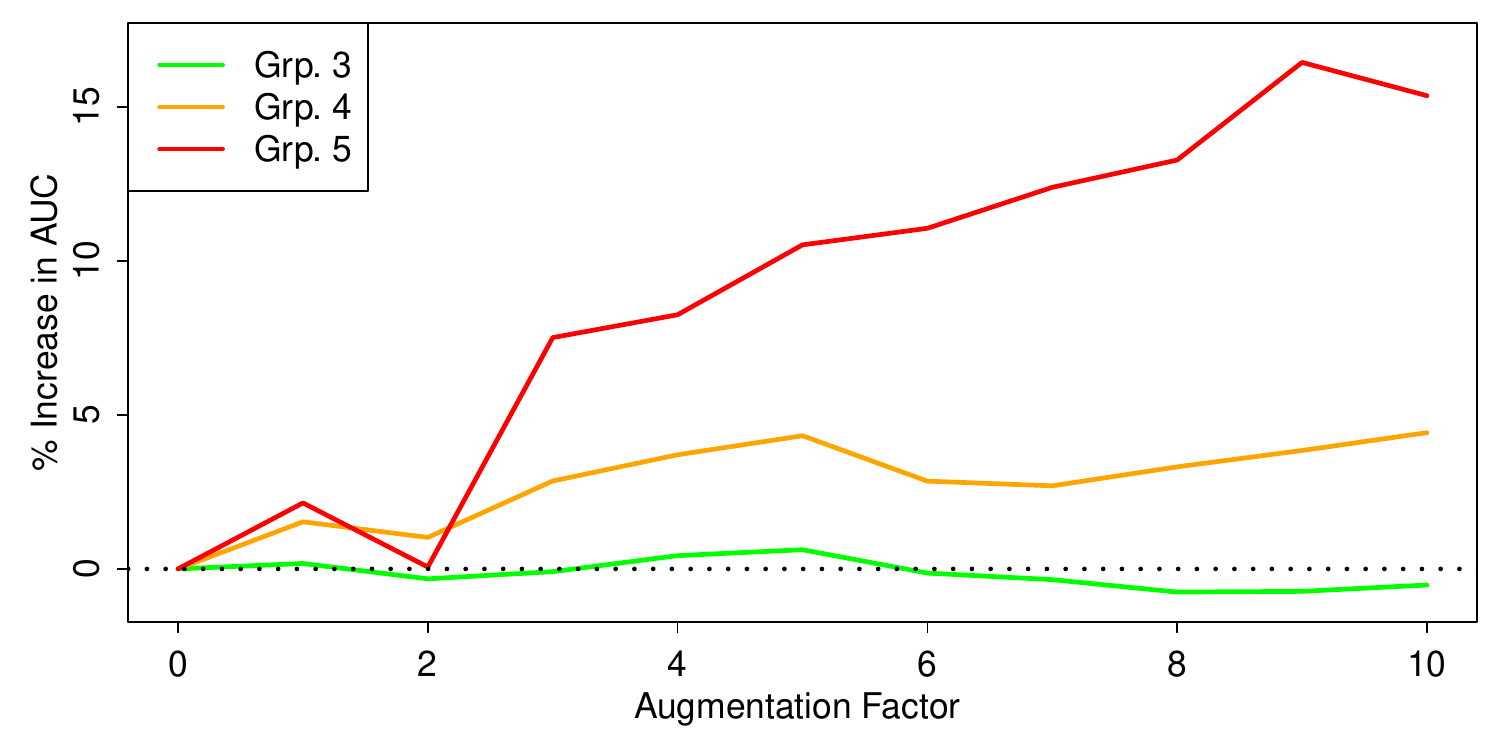}
 	\caption{Percent increase in AUC as a function of the augmentation factor. All groups are trained on the combined observed training sets from Groups~2--5 using a common augmentation factor in Groups~3--5, as plotted on horizontal axis. Zero corresponds to no augmentation and is the baseline against which increasing degrees of augmentation are compared.}
 	\label{fig43}
 \end{figure}

	
The current implementation of STACCATO requires the SN types for the validation {groups} in order to optimize the configuration of the augmented training {groups}. In a real application, the SN types for the test {groups} are unknown, and therefore the validation {groups are} not available. Simulation studies, however, could be used to optimize the configuration of the training sets before being rolled out on the actual test set. However, even in the absence of a definitive scheme for optimizing the augmentation scheme, Figure~\ref{fig43} demonstrates that augmenting {small training groups} is almost always beneficial. The figure depicts the effect of augmentation on AUC in Groups~3--5.  All three groups are trained using {(the original) Training} Groups~2--5 and using the same augmentation factor (from 0 to 10) in Groups~3--5. Group 3 is relatively insensitive to augmentation; this is also indicated in Table \ref{tab:optimal.composition}. Groups 4 and 5 both benefit from increasing levels of augmentation. In particular, the trend of the AUC as the augmentation factor increases is either positive or flat, with only local minor decrements. Thus, an augmentation factor of up to ten 
seems to be a safe choice {for small training groups}, even without optimizing the augmentation scheme using a validation {group}. 

We leave further investigation of both the tuning of STACCATO in the absence of a validation {group} and accounting for errors in  {or unavailable} redshift measurements to future work.

{\it Acknowledgements:} We thank David Stenning for helpful suggestions on the statistical methods developed in this paper and Bruce Bassett, Michelle Lochner and Rick Kessler for useful comments on a draft. This work was supported by Grant ST/N000838/1 from the Science and Technology Facilities Council (UK). RT was partially supported by an EPSRC ``Pathways to Impact'' grant. RT and DvD were supported by a Marie-Skodowska-Curie RISE (H2020-MSCA-RISE-2015-691164) Grant provided by the European Commission. 


\bibliographystyle{mnras}
\bibliography{ClassificationSNIa} 

\cleardoublepage
\appendix 

\section{Diffusion Maps}\label{app:diff.maps}

\subsection{Construction of the Diffusion Maps}\label{ssec:diff.maps}

Let $\Data_\band = \left\{ Y_i, \ldots, Y_N \right\}$ denote the normalised LC data in colour band $\band$ and define a metric $d_\band: \Data_\band \times \Data_\band \mapsto [0,\infty)$. The specific choice of $d_\band$ is introduced in Section \ref{subsec:light.curve.metric}. The metric is used to construct a Markov chain. The LCs form the state space and the probabilities of the Markov chain jumping from one LC to another are determined by the distance between the LCs under the metric, $d_\band$. Similar LCs are given high transition probabilities. The pairwise distances are transformed with a Gaussian kernel and  normalised so that the transition probabilities sum to one. Define the `weight' function, $w_\band:\Data_\band \times \Data_\band \mapsto (0,1]$, and the transition probabilities $p_\band(x,y)$ as
\begin{align}
w_\band(x,y) &= \exp \left( - \frac{d_\band(x,y)^2}{\varepsilon_\band} \right)\label{eq:w.func}\\ 
p_\band(x,y) &= \frac{w_\band(x,y)}{\sum_z w_\band(x,z)},\label{eq:transition}
\end{align}
where $\varepsilon_\band>0$ is a tuning parameter. Note that by applying (\ref{eq:w.func}) a distortion favouring `local jumps' in the Markov chain is introduced due to the exponential decay in (\ref{eq:w.func}). As a result, for many LC pairs $w_\band$ is small. For computational efficiency a sparsity parameter $\delta$ is introduced, such that if $w_\band(\cdot, \cdot) < \delta$ then it is set to zero; we use $\delta=10^{-5}$. With these definitions we can compute a sparse $N \times N$ transition matrix $P_\band$ giving the probabilities of jumping between all pairs of $\band$-band LC.

Define the diffusion distance between two LCs from the transition probabilities as
\begin{equation}\label{eq:diff.dis}
D_\band(x,y)^2 = \sum_{z \in \Data_\band} \frac{\left[ p_\band(x,z) - p_\band(y,z) \right]^2}{\phi_{0\band}(z)},
\end{equation}
(notice that $z$ here is a dummy index, not redshift) where $\phi_{0\band}(z)$ is the stationary distribution of the Markov chain,
\begin{equation}
\phi_{0\band}(x) = \frac{\sum_z w_\band(x,z)}{\sum_y \sum_z w_\band(y,z)}.
\end{equation}
Assuming that the Markov chain is irreducible, it follows from the construction of the chain that the stationary distribution exists and is unique \citep{coifman2006}. The diffusion distance in (\ref{eq:diff.dis}) measures the similarity between two LCs as the squared difference in probabilities of transitioning from $x$ to $z$ versus $y$ to $z$ in a weighted sum over all $z \in \Data_\band$. 

The diffusion distances can be calculated from the right eigenvalues and eigenvectors of the transition matrix $P_\band$, specifically,
\begin{eqnarray}\label{eq:eigen}
D_\band(x,y)^2 
&=& \sum_{j=1}^{N-1} \lambda_j^2 \left( \psi_j(x) - \psi_j(y) \right)^2 \\ &\approx& \sum_{j=1}^{m} \lambda_j^2 \left( \psi_j(x) - \psi_j(y) \right)^2, \label{eq:eigenapprox}
\end{eqnarray}
where $|\lambda_0| \geq |\lambda_1|\geq \cdots \geq |\lambda_{N-1}|$ are the eigenvalues sorted in descending order and $\psi_j$ are the corresponding eigenvectors normalised according to $\phi_{0\band}$, such that $\sum_x \psi_j^2 (x)\phi_{0\band}(x)=1$, with $\psi_j(x)$ equal to entry $x$ of eigenvector $j$. (To simplify notation we suppress the subscript $\band$ indicating the colour band for $\psi$ and $\lambda$.) The first eigenvector and eigenvalue has been left out of the sum in (\ref{eq:eigen}). This is because $\psi_0$ can be shown to be constant in its entries. Because the eigenvalues are in decreasing order, most of the diffusion distance can be explained by the first $m \ll N$ terms in the sum, hence the approximation in (\ref{eq:eigenapprox}). In our numerical results, we set $m$ to be the smallest value such that $\lambda_m < 0.05 \lambda_1$ (in (\ref{eq:eigenapprox})), with a maximum cap of 25.

Using (\ref{eq:eigenapprox}) the diffusion map, $\mathbf{\Psi}_\band: \Data_\band \mapsto \mathbb{R}^m$, is defined
\begin{equation}\label{eq:diff.map}
\mathbf{\Psi}_\band: x \mapsto \left[ \lambda_1 \psi_1(x), \lambda_2 \psi_2(x), \ldots, \lambda_m \psi_m(x) \right].
\end{equation}
From this mapping the approximate diffusion distances (\ref{eq:eigenapprox}) can be calculated as the Euclidean distance between the representation of the LCs in the diffusion space. For further details on diffusion maps, see e.g.  \citet{lafon2006} and \citet{coifman2006}.

\subsection{The Nystr\"{o}m Extension}\label{subsec:Nystrom}
\citet{Richards2012} apply the diffusion map to the entire dataset, including training and test sets. This approach results in what is called a `semi-supervised' classification algorithm, in that the unlabelled LCs are used to construct the representation of the LCs in $\mathbb{R}^m$. In contrast, we use only the training data  to construct the diffusion map. This has two benefits.  First, computing and storing the distance matrix $P_\band$ requires considerable computational resources and scales as $\mathcal{O} (N^2)$ where $N$ is the number of LCs. Hence computing the diffusion map on only about 7\% of the data is advantageous from a practical perspective. Second, in Section~\ref{sec:staccato} we show that computing the diffusion map on the training data only increases classification performance. However, this approach requires a method to map new LCs (i.e. the curves that need to be classified and were not included in the training set) into the diffusion space. The Nystr\"{o}m extension gives a simple method to do this.

The following outline of the Nystr\"{o}m extension is based on \citet{freeman2009}. Suppose there are $n$ $\band$-band LCs in the training set, and $k$ unlabelled $\band$-band LCs in the test set. The first step of the Nystr\"{o}m extension is to compute a $k \times n$ matrix $W_\band$ with entries equal to the normalized $w_\band(x,y)$ with $x$ representing a LC in the test set and $y$ a LC in the training set. Using the same $\varepsilon_\band$ parameter, the rows of $W_\band$ are normalised, as with $P_\band$, following (\ref{eq:transition}).

The eigenvectors used to construct the diffusion map in (\ref{eq:diff.map}) can be arranged in a $n \times m$ matrix $\Psi_\band$. The corresponding eigenvalues are used to construct an $m \times m$ diagonal matrix $\Lambda_\band$ with diagonal entry $i$ equal to $1/ \lambda_i$. Finally, the diffusion-space coordinates of the test set LCs are given by the rows of the $k \times m$ matrix built as follows:
\begin{equation}\label{eq:nystrom}
\Psi'_\band = W_\band \Psi_\band \Lambda_\band.
\end{equation}

\subsection{The Light Curve Metric}\label{subsec:light.curve.metric}
The mean squared difference,
\begin{equation}\label{eq:metric1}
d_\band(x,y) = \frac{1}{\tilde{t}_u^\band - \tilde{t}_l^\band + 1} \sum_{t = \tilde{t}_l^\band}^{\tilde{t}_u^\band} \left( \tilde{f}^\band_x (t) - \tilde{f}^\band_y (t) \right)^2, \hspace{3mm} x,y \in \Data,
\end{equation} 
is used as the LC metric, where $\tilde{t}_l^\band = \max \left( \tilde{t}^\band_{x,1}, \tilde{t}^\band_{y,1} \right) $ and $\tilde{t}_u^\band = \min \left(\tilde{t}^\band_{x,n_x^\band}, \tilde{t}^\band_{y,n_y^\band} \right)$ form the lower and upper bounds of the common time domain of the two LCs. When $\tilde{t}_u^\band < \tilde{t}_l^\band$, i.e., when there is no common domain for the two LCs, $d_\band(x,y)$ is set to one. (This is a large value relative to the scales of $\tilde{f}$.) Note that this does not necessary result in the two LCs being far apart in the diffusion space following (\ref{eq:diff.dis}).

\end{document}